\documentclass[prd, twocolumn, amsmath, amssymb, aps, floatfix, nofootinbib, superscriptaddress]{revtex4-2}

\usepackage{graphicx}
\usepackage{amsmath}
\usepackage{amssymb}
\usepackage{mathtools}
\usepackage{physics}
\usepackage{xcolor}

\newcommand{\eRPA}{\epsilon^{\mathrm{RPA}}}
\newcommand{\eLFE}{\epsilon^{\mathrm{LFE}}}
\newcommand{\enoLFE}{\epsilon^{\mathrm{noLFE}}}
\newcommand{\ame}{\alpha m_e}
\newcommand{\qbz}{\mathbf{q}_{\mathrm{1BZ}}}

\begin{document}

\title{All-electron dark matter-electron scattering with random-phase approximation dielectric screening and local field effects}

\author{Cyrus E. Dreyer}
\affiliation{Department of Physics and Astronomy, Stony Brook University, Stony Brook, NY 11794, USA}
\affiliation{Center for Computational Quantum Physics, Flatiron Institute, 162 Fifth Avenue, New York, NY 10010, USA}

\author{Rouven Essig}
\affiliation{C. N. Yang Institute for Theoretical Physics, Stony Brook University, Stony Brook, NY 11794, USA}

\author{Marivi Fernandez-Serra}
\affiliation{Department of Physics and Astronomy, Stony Brook University, Stony Brook, NY 11794, USA}
\affiliation{Institute for Advanced Computational Sciences, Stony Brook University, Stony Brook, NY 11794, USA}

\author{Megan Hott}
\email{Contact author: megan.hott@stonybrook.edu}
\affiliation{Department of Physics and Astronomy, Stony Brook University, Stony Brook, NY 11794, USA}
\affiliation{C. N. Yang Institute for Theoretical Physics, Stony Brook University, Stony Brook, NY 11794, USA}
\affiliation{Institute for Advanced Computational Sciences, Stony Brook University, Stony Brook, NY 11794, USA}

\author{Aman Singal}
\affiliation{Department of Physics and Astronomy, Stony Brook University, Stony Brook, NY 11794, USA}
\affiliation{C. N. Yang Institute for Theoretical Physics, Stony Brook University, Stony Brook, NY 11794, USA}
\affiliation{Institute for Advanced Computational Sciences, Stony Brook University, Stony Brook, NY 11794, USA}

\begin{abstract}
    Accurate predictions for dark matter-electron scattering in solids require an all-electron treatment together with a faithful description of dielectric screening beyond simple approximations. In particular, local field effects, arising from microscopic inhomogeneities of the electronic response, can significantly modify scattering rates across relevant momentum and energy scales. We present an all-electron framework for computing dark matter-electron scattering rates that incorporates dielectric screening at the random-phase approximation (RPA) level, including local field effects. Using crystalline silicon as a benchmark, we show that local field effects play an important role both at large momentum transfers, spanning multiple Brillouin zones, and at low momentum near the plasmon resonance. We compute electron recoil spectra and projected sensitivities for nonrelativistic halo dark matter and for boosted dark matter or other dark-sector particles, which are sensitive to the impact of local field effects in these high and low momentum regimes, respectively. We further present RPA dielectric functions including local field effects for Ge, GaAs, SiC, and diamond, enabling a systematic comparison across target materials. These developments are implemented in the open source code \textsc{QCDark2}.
\end{abstract}

\maketitle

\section{Introduction} \label{sec:intro}
    Several direct detection experiments searching for electron recoils in solid state materials~\cite{Essig2012} have been implemented or proposed in recent years, including SENSEI~\cite{Tiffenberg:2017aac,Crisler:2018gci,SENSEI:2019ibb,SENSEI:2020dpa,SENSEI:2021hcn,SENSEI1,SENSEI2,SENSEI:2025qvp}, DAMIC-M~\cite{DAMIC-M:2023gxo,DAMIC-M:2023hgj,DAMIC-M:2025ltz,DAMIC-M2025}, SuperCDMS~\cite{SuperCDMS:2018mne,SuperCDMS:2020ymb,SuperCDMS2025,SuperCDMS:2025dha}, EDELWEISS~\cite{EDELWEISS2020}, and Oscura~\cite{Oscura2022}. These experiments aim to detect light dark matter (DM), with masses ranging from $\mathcal{O}(\mathrm{keV})$ to $\mathcal{O}(\mathrm{GeV})$, which interact with electrons and possibly other standard model particles. Accurate calculations of DM-electron scattering rates and the expected spectrum are required to optimize DM searches and to understand the reach into parameter space that these experiments achieve.
    This involves the use of \textit{ab initio} methods to obtain the electronic structure of the detector material and to calculate its response to potential DM scattering events. 
    
    Density functional theory (DFT) is the standard tool for electronic structure calculations. When using it to calculate DM-electron scattering, there are several important theoretical aspects that need to be considered. 
    Since the DM interactions can transfer large amounts of momentum to the detector (several times the crystal momentum), an all-electron treatment of the electronic states is needed~\cite{EXCEED-DM,QCDark1}.
    The response of the detector material also involves dielectric screening, which is captured via the dielectric function~\cite{DarkELF,Hochberg2021}. The dielectric function can either be approximated analytically or calculated \textit{ab initio} at varying levels of theoretical accuracy.

    Existing implementations for calculating DM-electron scattering have taken different approaches to these two aspects of the electronic response (i.e., including all-electron effects and dielectric screening).
    The first \textit{ab initio} calculations were performed with \textsc{QEDark}~\cite{QEDark}, which is built on the plane-wave DFT software \textsc{Quantum ESPRESSO}~\cite{Giannozzi2017}. While plane-wave DFT codes are computationally efficient for solid-state applications, the use of pseudopotentials means that extensions are necessary to capture the response at high momentum \cite{EXCEED-DM,EXCEED-DM_code}. Additionally, in \textsc{QEDark}, the full effects of dielectric screening were neglected. 
    Results obtained with \textsc{DarkELF}~\cite{DarkELF,DarkELF_code}, which calculates the \textit{ab initio} random-phase approximation (RPA) dielectric function using \textsc{GPAW}~\cite{Mortensen2024}, have much more accurate screening for both low and intermediate momentum transfers. \textsc{DarkELF} provides tabulated dielectric functions for Si and Ge with and without local field effects. However, the dielectric functions are only calculated up to momentum transfers\footnote{We will write momentum in units of $\alpha m_e$, also known as inverse bohr or atomic units (a.u.). $1 \:\alpha m_e \approx 3.73\:\mathrm{keV}$. We will also set $\hbar = c = 1$ throughout.} of $q = 6\:\alpha m_e$, and some all-electron effects are neglected through the use of pseudopotentials.
    \textsc{EXCEED-DM}~\cite{EXCEED-DM,EXCEED-DM_code} does capture high-momentum effects through all-electron reconstruction via the projector-augmented wave approach~\cite{Blochl1994}; however, the RPA dielectric function at low momentum transfer is not modeled accurately. 
    The first version of \textsc{QCDark} (which we will refer to as \textsc{QCDark1})~\cite{QCDark1} implements all-electron DFT with \textsc{PySCF}~\cite{pyscf}; however, the dielectric screening was approximated with an analytic model. 

    The lack of full \textit{ab initio} screening coupled with an all-electron treatment means that none of the above implementations are simultaneously accurate for both low and high momentum transfers. While high momentum transfer is important for the typical nonrelativistic halo DM model, the low momentum region also becomes relevant for DM particles that have been boosted out of the halo velocity distribution to relativistic or semirelativistic speeds~\cite{Kurinsky2020,Knapen:2020aky,Essig2024,Liang2025}.
    Considering these boosted DM components can extend the sensitivity of detectors to lower DM masses, since these lighter particles would now have enough kinetic energy to excite electrons in the crystal~\cite{Emken2018,An2018,Emken:2021lgc,An:2021qdl,Emken2024,Bringmann2019,Ema2019,Liang2025,Alvey2019,Su2020,Jho2021,Das2021,Sun2025}. Moreover, dark-sector particles (such as DM or millicharged particles) can also be produced in accelerators (e.g., beam dumps) and scatter downstream in low-threshold detectors~\cite{SENSEI:2023gie,Oscura:2023qch,CONNIE:2024off,CONNIE:2024pwt,Essig:2024dpa}.  In all cases, the low-momentum dielectric response produces a distinct signal of several electron-hole pairs (due to the plasmon resonance) that can be easier to distinguish from background events than that of halo DM, which typically peaks toward a lower number of charge carriers~\cite{Essig2024,SENSEI2}.
    
    To obtain scattering rates that are accurate for boosted DM (or other dark-sector particles) as well as halo DM, we have updated \textsc{QCDark} to calculate the \textit{ab initio} RPA dielectric function including local field effects (LFEs) to use for both electronic transitions and screening at all momenta. With this new version of the code, \textsc{QCDark2}, we have computed RPA dielectric functions for Si, Ge, GaAs, SiC, and diamond.
    We find that for halo DM, LFEs reduce the projected reach by 20\%--50\% for masses above a few MeV across all materials. Further, while the total integrated spectrum is unchanged, LFEs do modify the shape of the expected electron recoil spectrum. 

    The rest of the paper is organized as follows. We will first review the general formalism for DM-electron scattering rates in Section~\ref{sec:Theory}, including the contributions from both halo DM and boosted DM. We then review the random-phase approximation and discuss how LFEs are included in the dielectric function. The computational implementation of the RPA dielectric function is briefly explained in Section~\ref{sec:Code}.
    In Section~\ref{sec:Results}, we analyze how LFEs alter the dielectric function and electron recoil spectra for different DM fluxes, compare electron recoil spectra with previously used analytic screening approximations, show projected sensitivities of all the materials listed above, and compare \textsc{QCDark2} to other DM-electron scattering codes. We include a further discussion of how this work fits into the condensed matter literature and recent generalized-response DM studies in Section~\ref{sec:Discussion}, and conclude in Section~\ref{sec:Conclusion}. The Appendixes contain a full derivation of the boosted DM-electron scattering rate (\ref{sec:scattering_rate_derivation}), the band structures and dynamic structure factors of all materials (\ref{sec:all_materials_details}), a derivation of the RPA dielectric function in the optical limit (\ref{sec:optical_limit}), and extra comparison of our RPA dielectric functions to those available in \textsc{DarkELF} (\ref{sec:DarkELF_comparison}).
\section{Theory} \label{sec:Theory}
\subsection{Dark sector model} \label{sec:Theory:Model}
    We will consider dark sector models containing a fermion $\chi$ of mass $m_{\chi}$ and a dark mediator. The dark mediator can be either a vector $V$ with mass $m_{V}$, or a scalar $\phi$ with mass $m_{\phi}$. The dark mediator couples to $\chi$ with strength $g_{\chi}$ and to the standard model (SM) electron with strength $g_{e}$. This results in one of the following effective Lagrangians:
    \begin{equation} \label{eq:lagrangian}
    \begin{split}
        \mathcal{L} & \supset - g_e V_{\mu} \Bar{e} \gamma^{\mu} e - g_{\chi} V_{\mu} \Bar{\chi} \gamma^{\mu} \chi \\ 
        \mathcal{L} & \supset - g_e \phi \Bar{e} e - g_{\chi} \phi \Bar{\chi} \chi \, .
    \end{split}
    \end{equation}
    
    A vector mediator can obtain its coupling to the electron through kinetic mixing with the standard model photon, hence this type of mediator is often called a dark photon. If $m_{V} = 0$, then $\mathcal{L}$ can instead be rotated into a basis where $\chi$ appears to be directly coupled to the photon with a small electric charge, called a millicharge. Our analysis in Section~\ref{sec:Results} will be constrained to models with scalar or vector mediators (which includes the dark photon and millicharged DM models), although there has been much recent work on generalized dark sector interactions with the SM, which we will briefly discuss in Section~\ref{sec:Discussion}.
    
\subsection{Dark matter-electron scattering rate} \label{sec:scattering_rate}
    The differential DM-electron scattering rate with respect to the transferred energy $\omega$ can be written as
    \begin{equation} \label{eq:general_rate}
        \frac{dR}{d\omega} = \int dv \frac{d\Phi}{dv} \frac{d\sigma}{d\omega} \ ,
    \end{equation}
    where $d\Phi/dv$ is the differential flux for DM particles with velocity $v$, and $d\sigma/d\omega$ is the differential DM-electron scattering cross section. Assuming the DM is described by the standard halo model, this flux is comprised of two components,
    \begin{equation} \label{eq:flux}
        \frac{d\Phi}{dv} =  \frac{d\Phi_{\mathrm{halo}}}{dv} +  \frac{d\Phi_{\mathrm{boosted}}}{dv} \ .
    \end{equation}
    The first term on the right is the (usual) flux of DM particles in the halo that can be described by a truncated Maxwell-Boltzmann velocity distribution. We can write this flux in terms of the velocity distribution, $f_{\chi}(\mathbf{v})$, to obtain the more-familiar form of the nonrelativistic DM-electron scattering rate~\cite{Trickle2020},
    \begin{eqnarray*}
        \frac{dR_{\mathrm{halo}}}{d\omega} 
        = \left( \frac{\rho_{\chi}}{m_{\chi}} \int d^3v f_{\chi}(\mathbf{v}) v \right)
        \frac{d\sigma}{d\omega} \\
        = \frac{1}{\rho_T} \frac{\rho_{\chi}}{m_{\chi}} \int d^3v f_{\chi}(\mathbf{v}) \frac{d\Gamma}{d\omega} \ ,
    \end{eqnarray*}
    where $\rho_{T}$ is the mass density of the detector, $\rho_{\chi}$ is the local mass density of DM, and $\Gamma$ is the rate for a given DM velocity and detector volume. We use the recommended parameters in~\cite{Baxter2021} to model the DM halo.
    
    The second term on the right in Eq.~(\ref{eq:flux}) is the flux of DM particles previously in the halo that have been boosted to velocity $v$ through some other interaction before reaching the detector. We will focus on solar-reflected DM (SRDM) where halo DM particles scatter off the solar plasma~\cite{Emken2018,An2018,Emken:2021lgc,An:2021qdl,Emken2024}, although there are many other possible sources of boosted DM, including interactions with cosmic rays in the interstellar medium~\cite{Bringmann2019,Ema2019,Liang2025} and the Earth's atmosphere~\cite{Alvey2019,Su2020}, supernova neutrinos~\cite{Jho2021,Das2021,Sun2025}, and heavy DM annihilation~\cite{Agashe2014}. This flux, while typically very small, can be important for low DM masses, as it contains DM particles that have large kinetic energies and thus can more easily leave a signal above the energy threshold of a detector. 

    For application to direct detection in semiconductors, we will calculate the differential cross section for DM scattering off the electron density in a solid-state material of total mass $m_T$, given by
    \begin{eqnarray} \label{eq:cross_section_3d}
        \frac{d\sigma}{d\omega} = 
        \frac{m_T}{\rho_T} \frac{(g_e g_{\chi})^2}{4 E_{\chi} v} 
        \int \frac{d^3 q}{(2\pi)^3}\frac{1}{E_{\chi}'}
        \delta(E_{\chi}' - E_{\chi} + \omega) \nonumber \\
        \times \frac{H_{A'}(\mathbf{q})}{(\omega^2 - |\mathbf{q}|^2 - m_{A'}^2)^2} S(\omega, \mathbf{q}) \ ,
    \end{eqnarray}
    where $\omega$ is the energy and $\mathbf{q}$ is the momentum transferred to the detector from a DM particle with initial energy $E_{\chi}$, final energy $E_{\chi}'$, and velocity $v$. We use $A'$ to refer to either a vector or scalar mediator. The form of function $H_{A'}(\mathbf{q})$ depends on the mediator type [see Appendix~\ref{sec:scattering_rate_derivation} for a full derivation of the cross section]:
    \begin{equation} \label{eq:H}
    \begin{split}
        H_{V}(\mathbf{q}) &= (E_{\chi} + E_{\chi}')^2 - |\mathbf{q}|^2 \\ 
        H_{\phi}(\mathbf{q}) &= 4 m_{\chi}^2 - (E_{\chi} - E_{\chi}')^2 + |\mathbf{q}|^2
    \end{split}
    \end{equation}
    The dynamic structure factor $S(\omega, \mathbf{q})$ captures all the microscopic details of the material response. $S(\omega, \mathbf{q})$ can be generalized to describe various interactions including nuclear recoils and phonon response; however, in this work we will focus on just the electronic response of the material. In this case, $S(\omega, \mathbf{q})$ can be written in terms of the loss function\footnote{See Section~\ref{sec:Discussion} for a discussion of the generalization of the loss function for highly-boosted DM.}, $\mathrm{Im} [-1 / \epsilon(\omega, \mathbf{q})]$, obtained from the longitudinal dielectric function, $\epsilon$:
    \begin{equation} \label{eq:dynamic_structure_factor}
        S(\omega, \mathbf{q}) = \frac{q^2}{2\pi\alpha} \mathrm{Im} \left( \frac{-1}{\epsilon(\omega, \mathbf{q})} \right) \ .
    \end{equation}

    While Eq.~(\ref{eq:cross_section_3d}) is valid for a general crystal~\cite{Boyd2023} and \textsc{QCDark2} calculates the full angular-dependent dynamic structure factor $S(\omega, \mathbf{q})$, we will make the approximation in this work that the crystal is isotropic and average over the angular directions of $\mathbf{q}$, to obtain $S(\omega, q)$, where $q=|\mathbf{q}|$.  This is a good approximation for crystals with cubic symmetry and allows us to easily integrate over the angular $\mathbf{q}$ coordinates using the delta function (see Eq.~(\ref{eq:delta_angle}) in Appendix~\ref{sec:scattering_rate_derivation} for more details). Writing $E_{\chi}'$ in terms of $E_{\chi}$ and parametrizing the interaction strength with the reference cross section,
    \begin{equation}
        \Bar{\sigma}_e = \frac{\mu_{\chi e}^2 (g_e g_{\chi})^2}{\pi (m_{A'}^2 + (\alpha m_e)^2)^2} \ ,
    \end{equation}
    where $\mu_{\chi e}$ is the DM-electron reduced mass, results in
    \begin{eqnarray} \label{eq:cross_section_iso}
        \frac{d\sigma}{d\omega} = 
        \frac{m_T}{\rho_T} \frac{\Bar{\sigma}_e}{16 \pi \mu_{\chi e}^2} \frac{1}{E_{\chi}(E_{\chi} - \omega)}
        \int dq \:qH_{A'}(q) \nonumber \\
        \times \frac{(m_A^2 + (\alpha m_e)^2)^2}{(\omega^2 - q^2 - m_{A'}^2)^2} S(\omega, q)
        \frac{\Theta(v - v_{\mathrm{min}})}{v^2} \ ,
    \end{eqnarray}
    where
    \begin{equation} \label{eq:v_min}
        v_{\mathrm{min}} = \frac{q}{2\gamma m_{\chi}} + \frac{\omega}{q} \ .
    \end{equation}
    Since the initial energy of the DM particle can be written in terms of its velocity ($E_{\chi} = \gamma m_{\chi} = m_{\chi}/\sqrt{1 - v^2}$), we can now plug Eq.~(\ref{eq:cross_section_iso}) into Eq.~(\ref{eq:general_rate}) to obtain the rate of DM-electron scattering in the detector material. 

    For nonrelativistic DM in the halo, we can assume the energy and momentum transfer are negligible compared to the DM mass, and Eq.~(\ref{eq:v_min}) ensures that $q \gg \omega$. In this limit, $H_{A'} \approx 4 m_{\chi}^2$ for both types of mediators, and Eq.~(\ref{eq:cross_section_iso}) simplifies to
    \begin{eqnarray} \label{eq:cross_section} 
        \left( \frac{d\sigma}{d\omega} \right)_{\mathrm{NR}}
        = \frac{m_T}{\rho_T} \frac{\Bar{\sigma}_e}{4 \pi \mu_{\chi e}^2}
        \int dq \: q |F_{\mathrm{DM}}(q)|^2
        S(\omega, q) \nonumber \\
        \times \frac{\Theta(v - v_{\mathrm{min}})}{v^2} \ ,
    \end{eqnarray}
    where we have introduced the dark matter form factor,
    \begin{equation} \label{eq:F_DM}
        F_{\mathrm{DM}}(q) = \frac{(\alpha m_e)^2 + m_{A'}^2}{q^2 + m_{A'}^2} \ .  
    \end{equation}
    We will later show results for the light mediator and heavy mediator limits of the form factor ($m_{A'} \ll \alpha m_e$ and $m_{A'} \gg \alpha m_e$, respectively):
    \begin{equation} \label{eq:F_DM_limits}
        F_{\mathrm{DM}}^{\mathrm{light}}(q) = \frac{(\alpha m_e)^2}{q^2}, \qquad 
        F_{\mathrm{DM}}^{\mathrm{heavy}}(q) = 1 \ .  
    \end{equation}

    We see that the only part of the calculation that varies depending on the detector material (besides the density) is the dynamic structure factor, which is derived from the dielectric function. We can make varying levels of approximation to calculate the dielectric function, ranging from simple analytical models to beyond-DFT approaches. One must be careful to use a dielectric function model that is accurate for the energy and momentum transfers relevant to the DM flux considered. In sections~\ref{sec:dielectric_function} and~\ref{sec:Code}, we describe our implementation of the random-phase approximation to calculate the dielectric function at both small and large momentum transfers, which are relevant for boosted DM and halo DM, respectively.

\subsection{Ionization yield}
    After a DM particle scatters off the electron density in a crystal, imparting some momentum and energy to excite an electron-hole pair, the electron and hole can each excite further electron-hole pairs. This ``secondary ionization" process continues until all electrons and holes have energies lower than the bandgap. The final number of excited electrons, $Q$, is the ionization yield. The number of DM-electron scattering events resulting in an ionization yield of $Q$ per unit detector mass and exposure time is
    \begin{equation} \label{eq:excitation_rate}
        \Delta R_Q = \int d\omega \frac{dR}{d\omega} p(\omega, Q) \ ,
    \end{equation}
    where $dR/d\omega$ is the scattering rate from Eq.~(\ref{eq:general_rate}) and $p(\omega, Q)$ is the probability to excite $Q$ electron-hole pairs with a transferred energy of $\omega$. A given $\omega$ does not always excite the same number of electron-hole pairs because the energies required for various valence-to-conduction band excitations are different due to the band structure; additionally, a fraction of the transferred energy is lost to phonons and does not contribute to electron-hole pair excitation. A phenomenological model of $p(\omega, Q)$ taking these two effects into account was made for silicon in~\cite{RK}, which we will use to compute $\Delta R_Q$ (also called the electron recoil spectrum) in Section~\ref{sec:Results}. To study the signals arising from halo DM and SRDM separately, we will split the total scattering rate according to the two components of the flux in Eq.~(\ref{eq:flux}). 

\subsection{RPA dielectric function} \label{sec:dielectric_function}
    We calculate the longitudinal dielectric function in the random-phase approximation
    (RPA) using Kohn-Sham (KS) wave functions and energies from DFT. Crystals are periodic in lattice vectors $\mathbf{R}$, which corresponds to a microscopic response in reciprocal space that depends on reciprocal lattice vectors $\mathbf{G}$. The RPA dielectric function can be written as~\cite{BohmPines1952,Ehrenreich1959,Adler1962,Wiser1963,Hybertsen1987}
    \begin{widetext}
    \begin{multline} \label{eq:RPA}
        \eRPA_{\mathbf{G}\mathbf{G'}}(\omega, \qbz) = \delta_{\mathbf{G}\mathbf{G'}} - \frac{4\pi \alpha}{V_{\mathrm{cell}}|\qbz+\mathbf{G}||\qbz+\mathbf{G'}|} 
        \\ 
        \times \sum_{i\mathbf{k}}\sum_{j\mathbf{k'}} (f_{i\mathbf{k}} - f_{j\mathbf{k'}}) \lim_{\eta \rightarrow 0} \frac{\mel{i\mathbf{k}}{e^{-i(\qbz+\mathbf{G})\cdot \mathbf{r}}}{j\mathbf{k'}}  \mel{j\mathbf{k'}}{e^{i(\qbz+\mathbf{G'})\cdot \mathbf{r}}}{i\mathbf{k}}}   {\omega - (\omega_{j\mathbf{k'}} - \omega_{i\mathbf{k}}) + i \mathrm{sgn}(\omega_{j\mathbf{k'}} - \omega_{i\mathbf{k}}) \eta} \ ,
    \end{multline}
\end{widetext}
    where $\ket{i\mathbf{k}}$ is the KS wave function of band $i$ at wave vector $\mathbf{k}$ in the first Brillouin zone (1BZ). This state has energy $\omega_{i\mathbf{k}}$ and occupation number $f_{i\mathbf{k}}$.
    We evaluate Eq.~(\ref{eq:RPA}) at temperature $T = 0$~K to simplify the occupation numbers $f_{i\mathbf{k}}$ to 0 for unoccupied states (conduction bands) or 2 for occupied states (valence bands). This is a good approximation for semiconductor detectors used in direct-detection experiments, which are held at low temperatures. 

    The energy denominator of Eq.~(\ref{eq:RPA}) can be divided into a real and imaginary part,
    \begin{equation*}
        \lim_{\eta \rightarrow 0} \frac{1}{\omega - \Tilde{\omega} + i \:\mathrm{sgn}(\Tilde{\omega}) \eta}
        = \mathcal{P} \left( \frac{1}{\omega - \Tilde{\omega}}\right) - i \pi \mathrm{sgn}(\Tilde{\omega}) \delta(\omega - \Tilde{\omega}) \ ,
    \end{equation*}
    where $\Tilde{\omega} = \omega_{j\mathbf{k'}} - \omega_{i\mathbf{k}}$. We will refer to these two terms as the principal value part and the spectral part; more precisely, we define the spectral part of the RPA dielectric function as 
    \begin{widetext}
    \begin{multline} \label{eq:eRPA_spectral}
        \epsilon^{S}_{\mathbf{G}\mathbf{G'}}(\omega, \qbz) = \frac{4\pi^2 \alpha}{V_{\mathrm{cell}}|\qbz+\mathbf{G}||\qbz+\mathbf{G'}|} 
        \sum_{i\mathbf{k}}\sum_{j\mathbf{k'}} 
        \mathrm{sgn}(\omega_{j\mathbf{k'}} - \omega_{i\mathbf{k}}) (f_{i\mathbf{k}} - f_{j\mathbf{k'}})
        \\
        \times \delta(\omega - (\omega_{j\mathbf{k'}} - \omega_{i\mathbf{k}}))
        \mel{i\mathbf{k}}{e^{-i(\qbz+\mathbf{G})\cdot \mathbf{r}}}{j\mathbf{k'}}  \mel{j\mathbf{k'}}{e^{i(\qbz+\mathbf{G'})\cdot \mathbf{r}}}{i\mathbf{k}} \ .
    \end{multline}
    \end{widetext}
    Explicitly calculating the principal value part to the same precision as the spectral part requires an impractical number of bands, so we instead directly calculate only the spectral part and obtain the principal value part through the Kramers-Kronig relations.\footnote{This is a common technique that is performed in other codes---the implementation in \textsc{GPAW}, for example, is very well documented~\cite{Shishkin2006,Miyake2000}.}

    While $\{\mathbf{k}\}$ theoretically cover all wav evectors in the 1BZ, we sample the KS wave functions at a finite number of initial state $\mathbf{k}$ and final state $\mathbf{k'}$ from Monkhorst-Pack grids during the DFT calculation. The overlaps, $\mel{i\mathbf{k}}{e^{-i(\qbz+\mathbf{G})\cdot \mathbf{r}}}{j\mathbf{k'}}$, will introduce a factor of $\delta_{\mathbf{k}+\qbz, \mathbf{k'}}$. Therefore, we evaluate Eq.~(\ref{eq:eRPA_spectral}) at all $\qbz \in \{\mathbf{k'}\} - \{\mathbf{k}\}$, where $\qbz$ is by definition in the 1BZ. The number of $\mathbf{G}$ included in the matrix is determined by the maximum momentum transfer $\mathbf{q}_{\mathrm{max}} = \qbz + \mathbf{G}_{\mathrm{max}}$ considered in the scattering rate calculations.
 
    The microscopic RPA dielectric function is then used to determine the finite-momentum dielectric function required in Eq.~(\ref{eq:dynamic_structure_factor}). We can write $\mathbf{q} = \qbz + \mathbf{G}$ for any momentum $\mathbf{q}$ transferred to the material, such that the finite-momentum dielectric function is
    \begin{equation} \label{eq:eps_LFE}
        \epsilon(\omega, \mathbf{q} = \qbz + \mathbf{G}) = \frac{1}{\left[ \eRPA_{\mathbf{G}\mathbf{G'}} (\omega, \qbz) \right]^{-1} \bigg|_{\mathbf{G'} = \mathbf{G}}} \ ,   
    \end{equation}
    where we first take the matrix inverse of $\eRPA_{\mathbf{G}\mathbf{G'}}$, then take the diagonal element corresponding to $\mathbf{G}$, and finally invert that single element.\footnote{Our definition of the finite-momentum dielectric function is equivalent to the macroscopic dielectric function found in the condensed matter literature, 
    \begin{equation*}
        \epsilon_M(\omega, \mathbf{q}) = \frac{1}{\left[ \eRPA_{\mathbf{G}\mathbf{G'}} (\omega, \mathbf{q}) \right]^{-1} \bigg|_{\mathbf{G'} = \mathbf{G} = 0}} \ ,
    \end{equation*}
    generalized to nonzero $\mathbf{G}$~\cite{InteractingElectrons}.}

    Performing the matrix inversion amounts to averaging over the microscopic inhomogeneities in the crystal, or LFEs. We have the option to neglect LFEs by setting all nondiagonal elements ($\mathbf{G'} \neq \mathbf{G}$) of the RPA dielectric function to 0, making the matrix inversion trivial. This approximation will be referred to as $\enoLFE$:
    \begin{equation} \label{eq:eps_noLFE}
        \epsilon^{\mathrm{noLFE}}(\omega, \mathbf{q} = \qbz + \mathbf{G}) = \eRPA_{\mathbf{G}\mathbf{G}}(\omega, \qbz) \ ,
    \end{equation}
    while the full calculation including all off-diagonal elements as in Eq.~(\ref{eq:eps_LFE}) will be called $\eLFE$.

    The reason one would neglect LFEs is that it is computationally intensive to calculate all the off-diagonal elements of Eq.~(\ref{eq:RPA}) when a large number of $\mathbf{G}$ are included in the matrix. Our application requires many 
    $\mathbf{G}$ to accommodate the high $q_{\mathrm{max}}$ needed for accurate calculation of the scattering rate.
    Programs including \textsc{GPAW}~\cite{Yan2011,Mortensen2024} and \textsc{turboEELS} in \textsc{Quantum ESPRESSO}~\cite{Timrov2015,Giannozzi2017} have the option to neglect LFEs, and it has been common practice to do so in DM-electron scattering calculations~\cite{QCDark1,EXCEED-DM_code}. We perform a detailed investigation of LFEs in silicon in Section~\ref{sec:Results:LFEs}, and show that they have a significant effect on predicted scattering rates in Section~\ref{sec:Results:LFE_Cutoff}. 

\section{Computational details} \label{sec:Code}
    We performed all DFT calculations with the python package \textsc{PySCF}~\cite{pyscf}, which uses localized Gaussian-type orbital basis sets. The basis set, exchange-correlation functional, and Monkhorst-Pack~\cite{MonkhorstPack1976} $k$-grid $\{\mathbf{k}\}$ must be specified to perform a self-consistent field (SCF) calculation in \textsc{PySCF}, which converges the wave function coefficients and energies at each $\mathbf{k} \in \{\mathbf{k}\}$. The number of states, or bands, at each $\mathbf{k}$ is the number of elements in the chosen basis set. This is different from a plane-wave basis, where the number of bands is determined by a maximum energy cutoff. Therefore, the chosen basis set must be large enough to include states up to the energy required to converge the total scattering rate (or the maximum energy to which an experiment is sensitive). We use the correlation-consistent polarized basis sets cc-pv(t+d)z for silicon~\cite{Dunning2001} and cc-pvtz for all other materials~\cite{Dunning1989,Woon1993,Wilson1999}, which ensures an accurate representation of the band structure within our maximum energy range of 50 eV (150 eV for diamond due to its much larger bandgap). We generate at least 64 conduction bands and use around 45 to cover the maximum transition energy, $\omega_{\mathrm{max}}$, to ensure convergence in the calculation of the dielectric function of each material. The band structures are shown in Appendix~\ref{sec:all_materials_details} (see Fig.~\ref{fig:bands_all_materials}).
    The Perdew-Burke-Ernzerhof (PBE) exchange-correlation functional~\cite{PBE1996} was used in all DFT calculations, and the sizes of the $k$-grids used for each material are listed in Table~\ref{tab:parameters}. We apply a scissor correction on the conduction band energies to correct for the low bandgaps predicted by PBE; the bandgaps before and after the scissor correction can also be found in Table~\ref{tab:parameters}.
    Finally, we specify a vector in reciprocal space, $\Delta \mathbf{q}$, by which the final state $k$-grid $\{\mathbf{k}'\}$ is shifted relative to $\{\mathbf{k}\}$. The magnitude of this shift determines the smallest $\qbz$ at which the dielectric function is calculated. The final KS states at $\{\mathbf{k}'\}$ are obtained with a non-SCF calculation in \textsc{PySCF} using the fully $k$-converged SCF density.

    \begin{table*} 
        \caption{\label{tab:parameters}DFT and dielectric function parameters used for all materials studied. These include the original bandgap predicted by the PBE functional and the corrected bandgap after applying a scissor correction (SC), or uniform shift of the conduction bands. The target bandgaps are taken from experimental data. The maximum momentum at which LFEs are included is $q_{\mathrm{LFE}}$, and we calculate the dielectric functions for $0 \leq q \leq q_{\mathrm{max}}$ and $0 \leq \omega \leq \omega_{\mathrm{max}}$. All materials below are in the diamond or zinc blende crystal structure.}
        \begin{ruledtabular}
        \begin{tabular}{cccccccc}

            Material & Basis & k-grid & \begin{tabular}{c}Bandgap\\(before SC)\\(eV)\end{tabular} & \begin{tabular}{c}Bandgap\\(after SC)\\(eV)\end{tabular} & \begin{tabular}{c}$q_{\mathrm{LFE}}$\\($\alpha m_e$)\end{tabular} & \begin{tabular}{c}$q_{\mathrm{max}}$\\($\alpha m_e$)\end{tabular} & \begin{tabular}{c}$\omega_{\mathrm{max}}$\\(eV)\end{tabular}\\ \hline
            Si       & cc-pv(t+d)z   & 8$\times$8$\times$8 & 0.63  & 1.1 \cite{Kittel_SolidState}  & 8     & 25 & 50 \\
            
            SiC     & cc-pvtz       & 8$\times$8$\times$8 & 1.35  & 2.36 \cite{Levinshtein2001} & 8     & 20 & 50 \\
            
            C        & cc-pvtz       & 8$\times$8$\times$8 & 4.12  & 5.5 \cite{Madelung1991} & 12    & 20 & 150 \\
            
            Ge       & cc-pvtz       & \begin{tabular}{c}10$\times$10$\times$10 (LFE)\\8$\times$8$\times$8 (noLFE)\end{tabular} 
            & 0.53  & 0.67 \cite{Madelung1991} & 7     & 20 & 50 \\
             
            GaAs     & cc-pvtz       & 8$\times$8$\times$8 & 1.20  & 1.42 \cite{Madelung1991} & 7     & 20 & 50 \\
        \end{tabular}
        \end{ruledtabular}
    \end{table*}

    We require the dielectric function to very high momentum, which means a large number of $\mathbf{G}$ are included in Eq.~(\ref{eq:RPA}) --- $\mathcal{O}(10^4)$ for momenta up to 20--25 $\ame$. Including LFEs drastically increases the computational cost, since the full $\epsilon_{\mathbf{G}\mathbf{G'}}$ matrix must be calculated and inverted, instead of only calculating the diagonal elements if LFEs are neglected. This motivated us to impose an LFE cutoff, $q_{\mathrm{LFE}}$, which is the maximum momentum at which the dielectric function includes LFEs. A composite finite-momentum dielectric function is formed for each material that concatenates $\epsilon^{\mathrm{LFE}}$ and $\epsilon^{\mathrm{noLFE}}$ at $q_{\mathrm{LFE}}$,
    \begin{equation}
        \epsilon^{\mathrm{comp}} = 
        \begin{cases}
            \epsilon^{\mathrm{LFE}} (\omega, q) & q \leq q_{\mathrm{LFE}} \\ 
            \epsilon^{\mathrm{noLFE}} (\omega, q) & q > q_{\mathrm{LFE}}
        \end{cases}
        \label{eq:e_comp}
    \end{equation}
    The LFE momentum cutoff is around $7-8 \:\alpha m_e$ for most materials, which includes $\mathcal{O}(10^3)$ $\mathbf{G}$ vectors. We show the convergence of electron recoil spectra and reach projections with $q_{\mathrm{LFE}}$ in Section~\ref{sec:Results}.

    The large number of $\mathbf{G}$ also requires us to bin the elements of $\epsilon^{\mathrm{LFE}}$ or $\epsilon^{\mathrm{noLFE}}$ in three-dimensional $\mathbf{q}$-space during the calculation. The bins are defined by selecting the number of angular bins, $(N_{\theta}, N_{\phi})$, to include in each spherical shell separated by a chosen magnitude $dq$. We use parameters $N_{\theta} = 9$, $N_{\phi} = 16$, and $dq = 0.02\:\ame$. The smallest $\qbz$ at which the dielectric function is calculated with Eq.~(\ref{eq:eRPA_spectral}) is the chosen shift of the final state $k$-grid. We choose this shift to be $\Delta \mathbf{q} \sim \Delta \mathbf{k}/2$, where $\Delta \mathbf{k}$ is the distance between $k$-points in the Monkhorst-Pack grid. We then use the optical ($\mathbf{q} \to 0$) limit of the dielectric function to calculate the bins closest to the origin---see Appendix~\ref{sec:optical_limit} for more details. 

    The code proceeds by calculating the microscopic dielectric function for all $\mathbf{G}$ according to Eq.~(\ref{eq:RPA}) for a given $\qbz$. Next, the finite-momentum dielectric function is obtained with Eq.~(\ref{eq:eps_LFE}) or (\ref{eq:eps_noLFE}) and binned before the dielectric function calculation for the next value of $\qbz$ is started. After this process is completed for all $\qbz$, any bins with missing entries are filled in by interpolation of the surrounding bins. The chosen $\{\mathbf{k'}\}$ shift\footnote{We initially calculated the high-momentum dielectric functions without LFEs with a shift of $|\Delta \mathbf{q}| = 0.01\:\alpha m_e$ before we realized that the optical limit is required for momenta this small. Since the dielectric function at high momentum is not affected by this smaller shift, we did not recalculate the dielectric functions without LFEs past the LFE cutoff, except for Si and diamond. Therefore, these two materials have consistent $\{\mathbf{k}'\}$ shifts of $\Delta \mathbf{q} = \Delta \mathbf{k}/2$ while all other materials have $\Delta \mathbf{q} = \Delta \mathbf{k}/2$ up to the LFE cutoff and $\Delta \mathbf{q} = 0.01 \times (1, 1, 1) \:\alpha m_e$ past the LFE cutoff.} 
    and use of the optical limit allow us to maximize the number of low-momentum bins that are calculated before interpolation.
    The three-dimensional binned dielectric function can be used in Eq.~(\ref{eq:cross_section_3d}) to calculate the general rate, or the angular average can be taken to compute the rate in the isotropic approximation in Eq.~(\ref{eq:cross_section_iso}). As discussed above, we make the isotropic approximation for all results in Section~\ref{sec:Results} due to the cubic symmetry of the materials in Table~\ref{tab:parameters}.

\section{Results} \label{sec:Results}
\subsection{Impacts of LFEs on the dynamic structure factor} \label{sec:Results:LFEs}

    \begin{figure*}
        \includegraphics[width=1\linewidth]{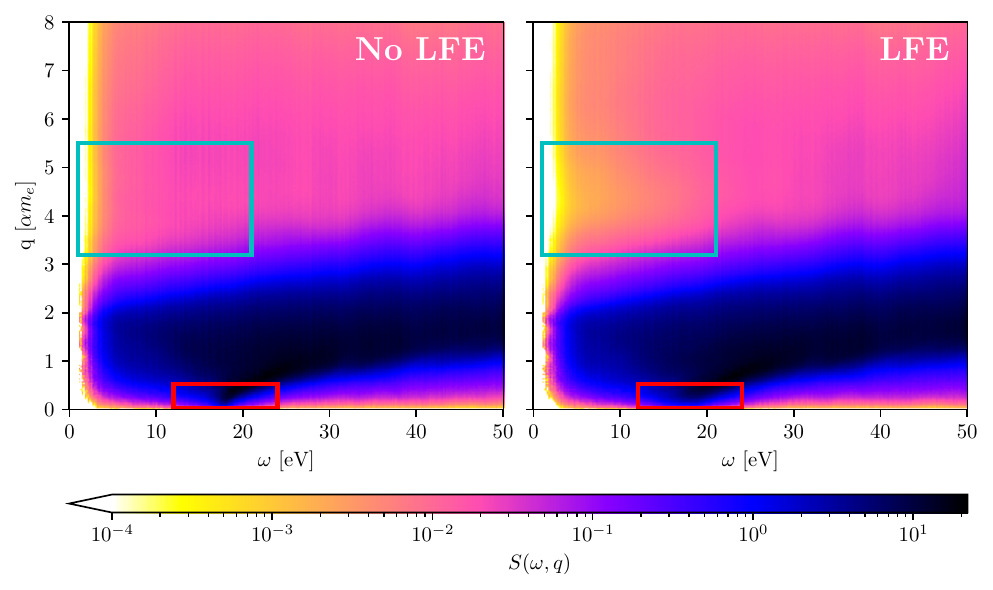}
        \caption{\label{fig:S_LFEs} Comparison of the dynamic structure factor (Eq.~(\ref{eq:dynamic_structure_factor}) averaged over angular $\mathbf{q}$ coordinates) for silicon calculated without local field effects (LFEs) on the left and with LFEs on the right. The red boxes at low momentum transfer and the cyan boxes at intermediate momentum transfer highlight the two areas where LFEs have the largest effect, as discussed further in the text.}
    \end{figure*}

\begin{figure*}
        \includegraphics[width=1\linewidth]{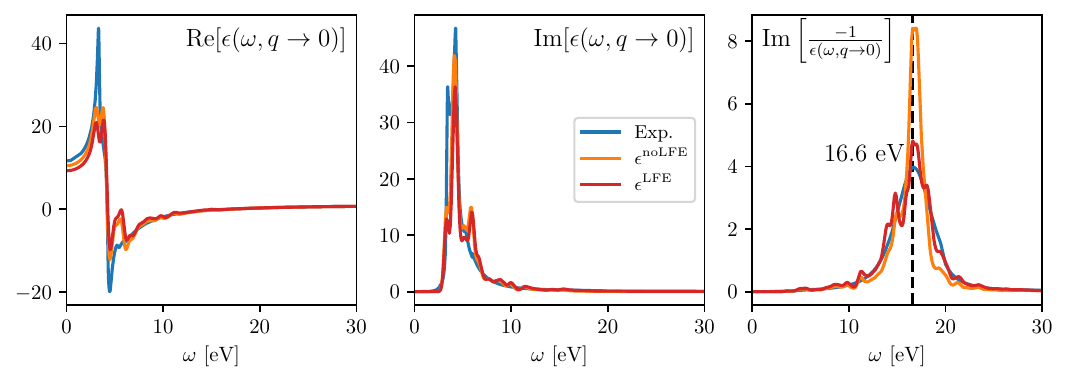}
        \caption{The dielectric function and loss function for Si calculated with \textsc{QCDark2} in the $q \rightarrow 0$ limit compared to experimental data from~\cite{Palik}. Gaussian smoothing with $\sigma_{\omega} = 0.2$~eV has been applied to the \textsc{QCDark2} curves for visual clarity. We see excellent agreement in the location of the plasmon peak at 16.6~eV between the RPA calculation and experiment.  The inclusion of LFEs broadens the plasmon peak, in better agreement with the data.}
        \label{fig:epsilon_data_comparison}
    \end{figure*}

\begin{figure*}
        \includegraphics[width=\linewidth]{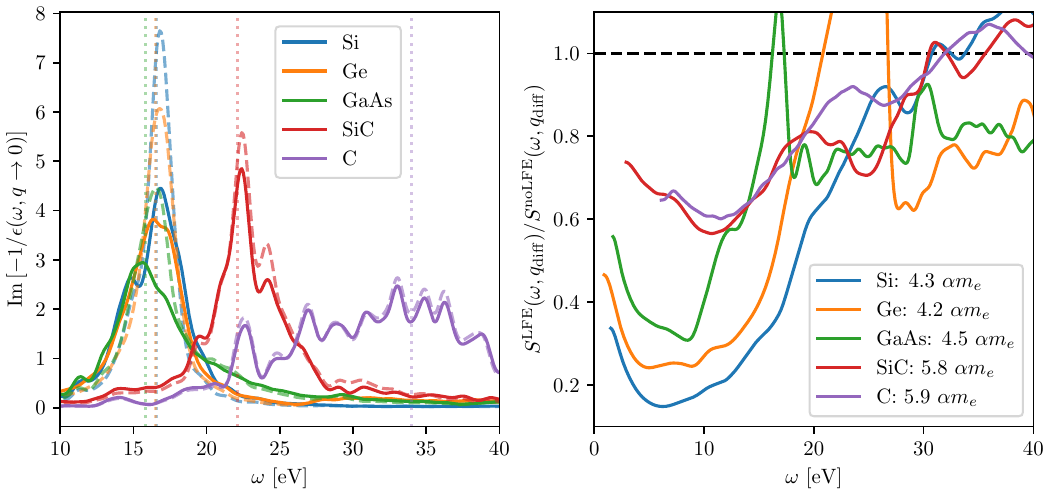}
        \caption{The left plot shows the loss function for all materials we consider, including Si, Ge, GaAs, SiC, and diamond (C), calculated with \textsc{QCDark2} in the $q \rightarrow 0$ limit. The dotted vertical lines correspond to the locations of the plasmon peak for each material taken from experimental data (Si:~\cite{Palik}, Ge:~\cite{Marton1967}, GaAs:~\cite{Brockt2000}, SiC:~\cite{Costantini2023}, diamond:~\cite{Serin1998,EELS_database}). The lighter dashed curves do not include LFEs, while the darker solid curves do include them. The right plot shows the ratio of the dynamic structure factor, $S$, with LFEs to $S$ without LFEs at $q_{\mathrm{diff}}$, the momentum where they differ the most for each material (which is the momentum indicated in the legend). Gaussian smoothing with $\sigma_{\omega} = 0.4$~eV has been applied to all curves for better visual clarity.}
        \label{fig:ELF_all_materials}
    \end{figure*}

    We first compare the isotropic dynamic structure factor of Si calculated with and without LFEs in Fig.~\ref{fig:S_LFEs}. There are two main areas in $(\omega, q)$ phase space that are significantly different when LFEs are included. The first occurs at low momentum transfers and is highlighted by the red box. A peak in the dynamic structure factor (as well as the loss function) can be observed in this region, which is due to collective plasmon excitations intrinsic to the material. It has been established that LFEs broaden the plasmon peak, which better matches experimental refraction index and electron energy loss spectroscopy data~\cite{Louie1975}.
    In Fig.~\ref{fig:epsilon_data_comparison}, we compare our computed dielectric function in the $q \rightarrow 0$ limit to the dielectric function at $q = 0$ obtained from experimental refractive index data tabulated in~\cite{Palik}. The broadening of the peak in the loss function is apparent with the inclusion of LFEs. The location of the peak at 16.6~eV is also consistent with the experimental data. The left plot of Fig.~\ref{fig:ELF_all_materials} shows the loss function in the $q \rightarrow 0$ limit for all the materials we consider in this study along with the expected peak locations taken from experimental data. Overall, we see very good agreement between experiment and our calculations.

\begin{figure*}
        \includegraphics[width=1\linewidth]{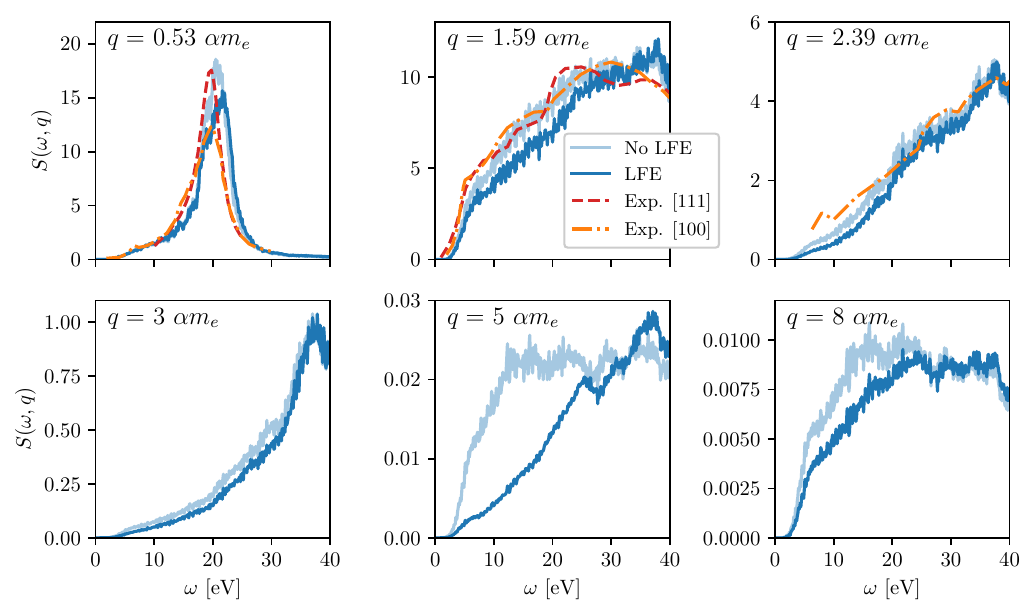}
        \caption{The top row shows the dynamic structure factor of Si with and without LFEs at intermediate momenta compared to experimental data from~\cite{Weissker2010}. The bottom row shows the dynamic structure taken at momentum slices below, inside, and above the cyan box in Fig.~\ref{fig:S_LFEs}, where we see significant differences between $S^{\mathrm{noLFE}}$ and $S^{\mathrm{LFE}}$. Above 3 $\alpha m_e$, LFEs greatly suppress the dynamic structure factor until $\sim$ 8 $\alpha m_e$.}
        \label{fig:high_mom_q_slices}
    \end{figure*}

    We compare the dynamic structure factor of Si with and without LFEs at finite momentum in Fig.~\ref{fig:high_mom_q_slices}. For momenta less than $\sim 3\:\alpha m_e$, we can compare our results to data from nonresonant inelastic x-ray scattering experiments. We include experimental data from~\cite{Weissker2010} along the $\mathbf{q} \parallel [111]$ and $\mathbf{q} \parallel [100]$ directions in the top row of Fig.~\ref{fig:high_mom_q_slices} for $0.53 \leq q \leq 2.39 \:\alpha m_e$. 

    Ref.~\cite{Weissker2010} calculated dielectric functions at the RPA level with and without LFEs, and additionally calculated a dielectric function in the adiabatic local-density approximation with time-dependent DFT (TDLDA)~\cite{RungeGross1984,GrossKohn1985}, which includes the exchange-correlation kernel that the RPA neglects (note that their TDLDA dielectric function still included LFEs). They found that the RPA dielectric function excluding LFEs reproduces the TDLDA and experimental results better than RPA with LFEs for this range of momenta. They attributed this to the cancellation of the exchange-correlation and LFE effects at the TDLDA level, particularly at energies $\omega \lesssim 20$~eV. However, the results in~\cite{Weissker2010} are only shown up to $q = 2.39\:\alpha m_e$, which is still below our second region of interest marked by the cyan box in Fig.~\ref{fig:S_LFEs}. Therefore, it is unknown whether this fortuitous agreement between experiment and RPA without LFEs continues at higher momenta. 
    Plotting our dynamic structure factors with and without LFEs at high momenta in the bottom row of Fig.~\ref{fig:high_mom_q_slices}, we find that they converge around $q \approx 3\:\alpha m_e$ before LFEs begin to greatly suppress the dynamic structure factor at higher momenta,

    The rest of the materials considered show a similar suppression when LFEs are included at high momentum. The right plot of Fig.~\ref{fig:ELF_all_materials} shows the ratio of the dynamic structure factors with and without LFEs at the momenta where they differ the most, $q_{\mathrm{diff}}$, which falls between 4--6 $\alpha m_e$ for each material. LFEs have the greatest effect on Si, Ge, and GaAs,\footnote{The impact of LFEs (and excitonic effects) on the dynamic structure factor in GaAs is studied in the context of DM detection in~\cite{Taufertshofer2025}, however they only perform calculations at momentum transfers $q < 4\:\alpha m_e$.} with the dynamic structure factor of Si being reduced to 15\% of its value without LFEs at low energies. These materials also have the largest reduction in plasmon peak height in the left plot of Fig.~\ref{fig:ELF_all_materials}, which are all located around 16 eV. The plasmon peaks of SiC and diamond are located at higher energies and do not have a significant change in height when LFEs are incorporated.

\subsection{Impacts of LFEs on electron recoil spectra} \label{sec:Results:LFE_Cutoff}
    We defined the composite dielectric function, which includes LFEs only up to momentum $q_{\mathrm{LFE}}$, by concatenating $\epsilon^{\mathrm{LFE}}$ and $\epsilon^{\mathrm{noLFE}}$ in Section~\ref{sec:Code} [Eq.~(\ref{eq:e_comp})]. In this section, we will analyze how the choice of $q_{\mathrm{LFE}}$ affects the predicted response in Si for both nonrelativistic halo DM and semirelativistic solar-reflected DM (SRDM). 

    The maximum velocity of the DM flux, $v_{\mathrm{max}}$, determines the region in $(\omega,q)$ phase space of the dynamic structure factor that is relevant to the scattering rate calculation. 
    When the cross section containing the velocity step function in Eq.~(\ref{eq:cross_section_iso}) is integrated with the flux to obtain the scattering rate, the constraints imposed by the step function become equivalent to the following kinematic limits on $q$ (see also Appendix~\ref{sec:scattering_rate_derivation}):
    \begin{align} \label{eq:q_lim}
        \begin{split}
            q_{\mathrm{min}} &= \gamma m_{\chi} v - \sqrt{(\gamma m_{\chi} - \omega)^2 - m_{\chi}^2} \\
            q_{\mathrm{max}} &= \gamma m_{\chi} v + \sqrt{(\gamma m_{\chi} - \omega)^2 - m_{\chi}^2} \ .
        \end{split}
    \end{align}
    The first example we will consider is halo DM with $m_{\chi} = 1$ GeV. In this case, the maximum velocity of the DM is $v_{\mathrm{max}} = v_{\mathrm{escape}} + v_{\mathrm{earth}} = 0.0026$, where $v_{\mathrm{escape}}$ is the Galactic escape speed and $v_{\mathrm{earth}}$ is the average speed of the Earth relative to the Galactic rest frame.\footnote{Using the recommended parameters in~\cite{Baxter2021}, $v_{\mathrm{escape}} = 544 \: \mathrm{km/s} = 0.0018$, $v_{\mathrm{earth}} = 250.2 \: \mathrm{km/s} = 0.0008$, and the local DM density is $\rho_{\chi} = 0.3$ $\mathrm{GeV}/\mathrm{cm}^3$.}
    For nonrelativistic velocities and large DM masses, the lower bound on the momentum transfer is almost mass independent and the upper bound greatly exceeds momenta where the dynamic structure factor contributes appreciably ($q_{\mathrm{max}} \sim 100\:\alpha m_e$).
    We will use a reference cross section of $\Bar{\sigma}_e = 10^{-39}\:\mathrm{cm}^2$ for the halo DM example to be consistent with the analysis in~\cite{QCDark1}.

    We will also consider SRDM fluxes with two different sets of parameters, one with $m_{\chi} = 50$ keV and reference cross section $\Bar{\sigma}_e = 10^{-38}\:\mathrm{cm}^2$, and the other with $m_{\chi} = 0.5$ MeV and $\Bar{\sigma}_e = 10^{-37}\:\mathrm{cm}^2$. Both SRDM examples assume a massless dark mediator. The maximum velocities were extracted from the start of the high-velocity tail of the flux distributions simulated in~\cite{Emken2024}: the 50 keV case has $v_{\mathrm{max}} = 0.06$ and the 0.5 MeV case has $v_{\mathrm{max}} = 0.02$. The reference cross sections were chosen such that both examples could be produced via a freeze-in mechanism (see calculations/benchmark curves in~\cite{Essig2012,Chu:2011be,QEDark,Dvorkin:2019zdi,Emken2024}).  

    \begin{figure*}
        \includegraphics[width=1\linewidth]{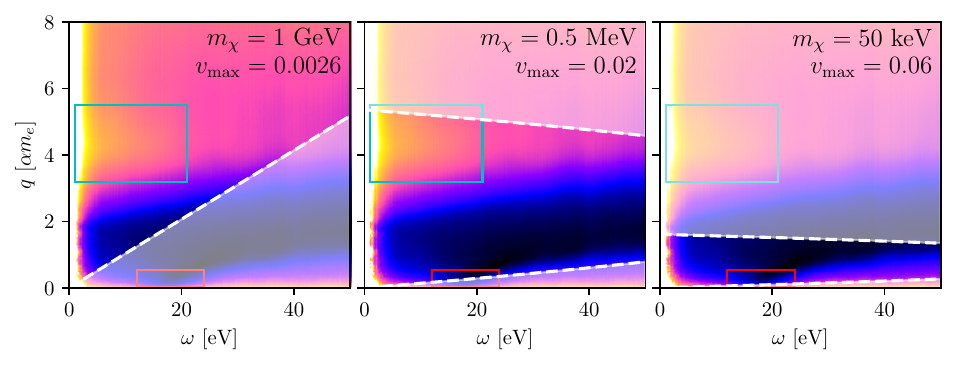}
        \caption{Kinematic limits on $(\omega, q)$ (white dashed curves) for three DM fluxes. The limits are overlaid on the dynamic structure factor (with LFEs) of Si from Fig.~\ref{fig:S_LFEs}, showing the two regions most affected by LFEs in the red and cyan boxes. The regions excluded by the kinematic limits are shaded. The left plot shows the lower kinematic limit on halo DM for a DM mass of $m_{\chi} = 1$ GeV, which excludes the low-momentum plasmon region but does include the intermediate-momentum region. The middle and right plots show the lower and upper kinematic bounds on SRDM with masses of $m_{\chi} = 0.5$~MeV and $m_{\chi} = 50$~keV, respectively. The $m_{\chi} = 0.5$~MeV case includes contributions from both regions affected by LFEs, while the $m_{\chi} = 50$~keV case is only impacted by the plasmon region.
        The maximum velocities of $v = 0.02$ and $v = 0.06$ for the two SRDM examples are determined by the fluxes simulated in~\cite{Emken2024} for a massless dark mediator.}
        \label{fig:S_DM_limits}
    \end{figure*}
    
    The regions of interest for all three fluxes (the halo DM flux and the two SRDM fluxes) in the $q$ versus $\omega$ parameter space, which are defined by the kinematic limits, are shown in Fig.~\ref{fig:S_DM_limits}. For halo DM, the kinematic limits exclude the low-momentum plasmon region of the Si dynamic structure factor, but include the intermediate-momentum region where LFEs are also relevant. Conversely, the 50~keV SRDM limits include the plasmon region but exclude the intermediate-momentum region. The 0.5~MeV SRDM limits include contributions from both regions. These examples were chosen to give a complete picture of how LFEs affect the electron recoil spectrum, and to emphasize the importance of having an accurate dielectric function for the entire $(\omega, q)$ space. The chosen fluxes also produce equivalent results for dark sectors with either vector or scalar mediators (see Section~\ref{sec:Theory:Model}), so all results in this section and the following section apply to both models.

    \begin{figure*}
        \includegraphics[width=1\linewidth]{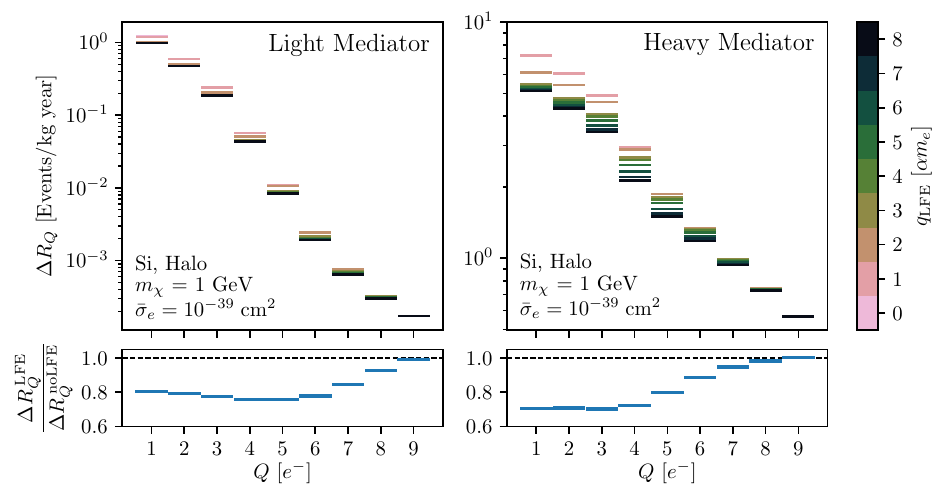}
        \caption{Electron recoil spectra in Si for halo DM with the ionization yield model from~\cite{RK} at 100~K, assuming a light mediator on the left and a heavy mediator on the right. The light mediator case converges quickly with LFE momentum cutoff, $q_{\mathrm{LFE}}$, since high momentum modes are suppressed by a factor of $q^{-4}$ compared to the heavy mediator case. The bottom panel shows the ratio of the rates when including LFEs (with $q_{\rm LFE}=8 \alpha m_e$) to not including them.} 
        \label{fig:composite_epsilon_recoil_spectrum}
    \end{figure*}

    Figure~\ref{fig:composite_epsilon_recoil_spectrum} shows how the electron recoil spectrum calculated with Eq.~(\ref{eq:excitation_rate}) for DM in the standard halo model is affected by the LFE momentum cutoff, $q_{\mathrm{LFE}}$, for the two limiting cases of the dark matter form factor as defined in Eq.~(\ref{eq:F_DM_limits}). The light mediator case converges quickly with $q_{\mathrm{LFE}}$, since high momentum modes are suppressed by a factor of $q^{-4}$ compared to the heavy mediator case. The heavy mediator case accesses more of the dynamic structure factor in and above the cyan rectangle in Fig.~\ref{fig:S_LFEs} where LFEs significantly reduce its magnitude. This causes scattering rates to be reduced by about 30\% for ionization yields of 1--4 electron-hole pairs.

    \begin{figure*}
        \includegraphics[width=1\linewidth]{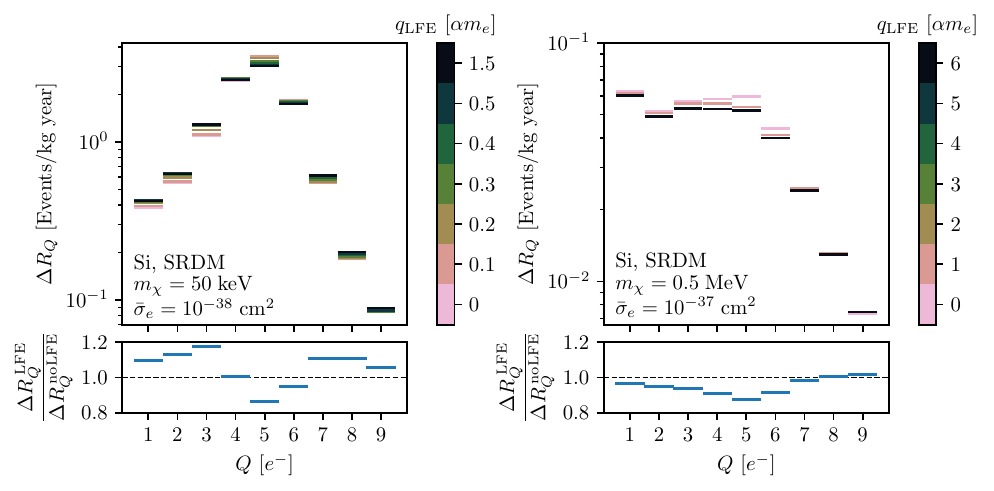}
        \caption{The electron recoil spectra in Si for two examples of solar-reflected DM fluxes assuming a massless dark mediator. The left plot corresponds to a DM mass of 50 keV while the right plot corresponds to 0.5 MeV. The reference cross sections are chosen to make both examples compatible with the benchmark freeze-in model~\cite{Essig2012,Chu:2011be,QEDark,Dvorkin:2019zdi,Emken2024}. The LFE momentum cutoff is increased over the relevant $(\omega, q)$ phase space in Fig.~\ref{fig:S_DM_limits} for each example. The broadening of the plasmon peak for higher $q_{\rm LFE}$ clearly redistributes some power from the peak at $Q = 5$ to $Q = $ 1--3 and 7--9 in the left plot, while the effects from including LFEs in the plasmon and intermediate-momentum regions compete in the right plot. The bottom panel shows the ratio of the rates when fully  including LFEs to not including them.} 
        \label{fig:SRDM_recoil_spectrum_lfe}
    \end{figure*}

    The electron recoil spectra with varying $q_{\mathrm{LFE}}$ for the two SRDM examples (both with a massless mediator) are shown in Fig.~\ref{fig:SRDM_recoil_spectrum_lfe}. Focusing first on the 50~keV case, the main feature of the spectrum is the enhancement in $\Delta R_Q$ for yields of 4--6 electron-hole pairs compared to the monotonic decrease in the event rate for halo DM. This enhancement is due to the inclusion of the plasmon peak at 16--18 eV in the cross section.\footnote{The mean energy per pair in silicon has been measured to be around 3.6--3.75~eV~\cite{Sellin2006,Rodrigues:2020xpt}, which corresponds to an average yield of 4.5--4.7 electrons for an energy transfer of 17 eV.} Since direct detection experiments see the most background signal in lower $Q$ bins (particularly $Q = 1$), higher sensitivity can be achieved for a signal peaked at these larger $Q$ bins~\cite{SENSEI2,Essig2024}. 
    The effects of the broadening of the plasmon peak by LFEs are also apparent in the 50~keV SRDM spectrum. This broadening causes power to be redistributed from the peak at 16.6~eV out to both lower and higher energies, which can be seen in the loss function in the third plot of Fig.~\ref{fig:epsilon_data_comparison} at $q \rightarrow 0$. This causes an increase in $\Delta R_Q$ for $Q = $ 1--3 and 7--9, while the peak at $Q = 5$ decreases as $q_{\mathrm{LFE}}$ is increased from 0 to $0.5\;\ame$. 

    The spectrum of the 0.5 MeV case is an intermediate example between nonrelativistic halo DM and light, (semi)relativistic SRDM where both the plasmon and high-momentum regions of the dynamic structure factor are relevant. We can still see the decrease in the middle bins as LFEs are included up to 1 $\alpha m_e$, but the increase in $\Delta R_Q$ for $Q = $ 1--3 and 7--9 caused by the broadening of the plasmon peak has been canceled out by the overall decrease in rates as LFEs are included at higher momenta. The flat spectrum is also a unique signal that could be easier to distinguish from backgrounds than that of halo DM. 

\subsection{Impacts of different screening approximations on electron recoil spectra} \label{sec:Results:Screening}
    \begin{figure*}
        \includegraphics[width=\linewidth]{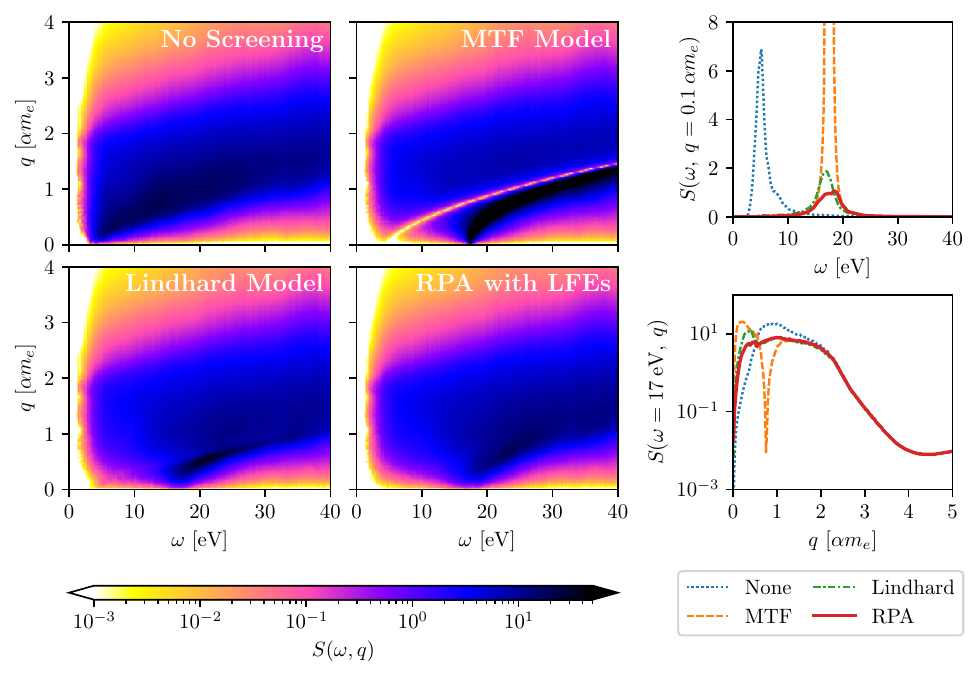}
        \caption{The left panel shows the dynamic structure factor of Si with different dielectric function approximations used for screening, i.e., $S(\omega, q) = \frac{q^2}{2\pi\alpha} \mathrm{Im}[\epsilon^{\mathrm{comp}}(\omega, q)] \:/\: |\epsilon^{\mathrm{scr}}(\omega, q)|^2$. Top left: $\epsilon^{\mathrm{scr}} = 1$ for no screening; top right: $\epsilon^{\mathrm{scr}} = \epsilon^{\mathrm{MTF}}$ [Eq.~(\ref{eq:MTF})]; bottom left: $\epsilon^{\mathrm{scr}} = \epsilon^{\mathrm{Lindhard}}$ [Eq.~(\ref{eq:Lindhard})]; bottom right: $\epsilon^{\mathrm{scr}} = \eLFE$. 
        The right panel shows slices of the dynamic structure factor at a constant momentum of $q = 0.1\:\alpha m_e$ in the upper plot and at constant energy $\omega = 17$ eV in the lower plot. These examples highlight the importance of accurate screening at low momentum and show the eventual convergence of all screening models at large enough momentum.}
        \label{fig:S_screening}
    \end{figure*}

    We can write the loss function (or equivalently the dynamic structure factor) in terms of the real and imaginary parts of the dielectric function:
    \begin{equation}
        \mathrm{Im}\left( \frac{-1}{\epsilon(\omega, q)} \right) = \frac{\mathrm{Im}[\epsilon(\omega, q)]}{\mathrm{Re}[\epsilon(\omega, q)]^2 + \mathrm{Im}[\epsilon(\omega, q)]^2} \ ,
    \end{equation}
    where the denominator represents the effects of dielectric screening in the medium. In previous works~\cite{QCDark1,EXCEED-DM}, only the imaginary part of the RPA dielectric function, $\mathrm{Im}[\epsilon(\omega, q)]$, was calculated to use as the numerator, while different approximations of the dielectric function, which we will call $\epsilon^{\mathrm{scr}}$, were used to obtain the magnitude of the dielectric function for the denominator. This was done to avoid the calculation of the real part of the RPA dielectric function. The loss function for a given screening approximation can be written as
    \begin{equation} \label{eq:ELF_screening}
        \mathrm{Im}\left( \frac{-1}{\epsilon(\omega, q)} \right)_{\mathrm{scr}} = \frac{\mathrm{Im}[\epsilon(\omega, q)]}{|\epsilon^{\mathrm{scr}}(\omega, q)|^2} \ .
    \end{equation}

    In this section, we will compare electron recoil spectra in Si calculated with no screening, i.e. $\epsilon^{\mathrm{scr}} = 1$, full RPA screening using the composite dielectric function with LFEs, and two approximations used previously for \textsc{QCDark1} analyses: the analytic model taken from~\cite{Cappellini1993}, which we will call the modified Thomas-Fermi (MTF) model, and the Lindhard approximation~\cite{Lindhard1953,Vos2025}. 
    All models will use $\mathrm{Im}[\epsilon^{\mathrm{comp}}]$ with $q_{\mathrm{LFE}} = 8\:\alpha m_e$ [see Eq.~(\ref{eq:e_comp}) for the definition of $\epsilon^{\mathrm{comp}}$] as the numerator of Eq.~(\ref{eq:ELF_screening}). 

    The MTF model was used in the first version of \textsc{EXCEED-DM}~\cite{EXCEED-DM_code} as well as \textsc{QCDark1}~\cite{QCDark1}. The material-dependent and fitted parameters included in this model are discussed in~\cite{Cappellini1993}; we take their best-fit parameters for Si as $\epsilon_0 = 11.3$, $\tau = 1.563$, $\omega_p = 16.6$ eV, and $q_{\mathrm{TF}} = 4.13$ keV to obtain
    \begin{multline} \label{eq:MTF}
        \epsilon^{\mathrm{MTF}}(\omega, q)
        = 1 \; + \\
        \left[ \frac{1}{\epsilon_0 - 1} + \tau \left( \frac{q}{q_{\mathrm{TF}}}\right)^2 + \frac{q^4}{4 m_e^2 \omega_p^2} - \left( \frac{\omega}{\omega_p} \right)^2 \right]^{-1} \ .
    \end{multline}

    The Lindhard model can be derived from the RPA dielectric function in Eq.~(\ref{eq:RPA}) by plugging in the electronic states for a homogeneous electron gas of density $n_e$. This greatly simplifies the wave functions and all overlaps in the numerator reduce to 0 or 1. Further algebra can be performed to write the Lindhard dielectric function as~\cite{Lindhard1953,Vos2025}
    \begin{equation} \label{eq:Lindhard}
        \epsilon^{\mathrm{Lindhard}}(\omega, q) = 1 + \frac{3 \omega_p^2}{q^2v_F^2} \left( \frac{1}{2} + F_{+} + F_{-} \right) \ ,
    \end{equation}
    where
    \begin{equation*}
        F_{\pm} = \frac{q_F}{4q} (1 - Q_{\pm}^2) \mathrm{Log}\left[ \frac{Q_{\pm} + 1}{Q_{\pm} - 1}\right]
    \end{equation*}
    and
    \begin{equation*}
        Q_{\pm} = \frac{q}{2 q_F} \pm \frac{\omega}{q v_F} \pm i \frac{f_p \omega_p}{q v_F} \ .   
    \end{equation*}
    Often the Lindhard approximation takes the imaginary part of $Q_{\pm}$ to 0 (as was originally done by Lindhard in~\cite{Lindhard1953}); however, we choose a finite $f_p$ to match $\mathrm{Im}[-1/\epsilon^{\mathrm{Lindhard}}(\omega, q = 0)]$ to experimental results, since we are interested in the plasmon region at finite momentum. For Si, we choose $f_p = 0.1$. The other parameters in the Lindhard model are derived from the average electron density in Si: $q_F = (3\pi^2 n_e)^{1/3}$, $v_F = q_F / m_e$, and $\omega_p = \sqrt{4\pi \alpha n_e / m_e} = 16.6$~eV.

    First we show the dynamic structure factor of Si with the above screening models in the left panel of Fig.~\ref{fig:S_screening}. All three screening models greatly suppress the low-energy absorption peaks of the numerator, as can be seen in the upper plot of the right panel where a slice of the dynamic structure factor is taken at constant momentum $q = 0.1\:\alpha m_e$. 
    MTF screening has been used in studies of halo DM~\cite{QCDark1,EXCEED-DM,DAMIC-M2025,SENSEI1,SENSEI2} since it accurately captures the magnitude of the dielectric function in the large-$q$ limit; however, it should not be used when considering boosted DM fluxes due to the singularity at low-$q$ and extremely high magnitude of the plasmon peak. These features can be seen in the right panel of Fig.~\ref{fig:S_screening}: the lower plot shows that the dynamic structure factor at constant energy $\omega = 17$~eV screened with the MTF model agrees well with RPA screening past the singularity at $q \approx 1\:\alpha m_e$, and the upper plot shows the inaccurate plasmon peak at low momentum.
    The Lindhard model was used previously to study boosted dark matter~\cite{Essig2024,Emken2024} (and low-energy Compton scattering~\cite{Essig2024_Compton}), since it models the plasmon region more accurately than the MTF model (as can be seen in the upper right plot of Fig.~\ref{fig:S_screening}); however, it still cannot capture material-specific features such as interband matrix elements and LFEs that become important in Si at larger momentum transfer.
    Additionally, constructing the loss function using $\mathrm{Im}[\epsilon]$ from one approximation and $|\epsilon|^2$ from another is of course not fully self-consistent because the real and imaginary parts of $\epsilon$ are linked by causality. By computing the full complex RPA dielectric function, we incorporate screening consistently across $(\omega, q)$, and we quantify the residual model dependence by comparing scattering rates computed with each of these screening models.

    \begin{figure*}
        \centering
        \includegraphics[width=0.9\linewidth]{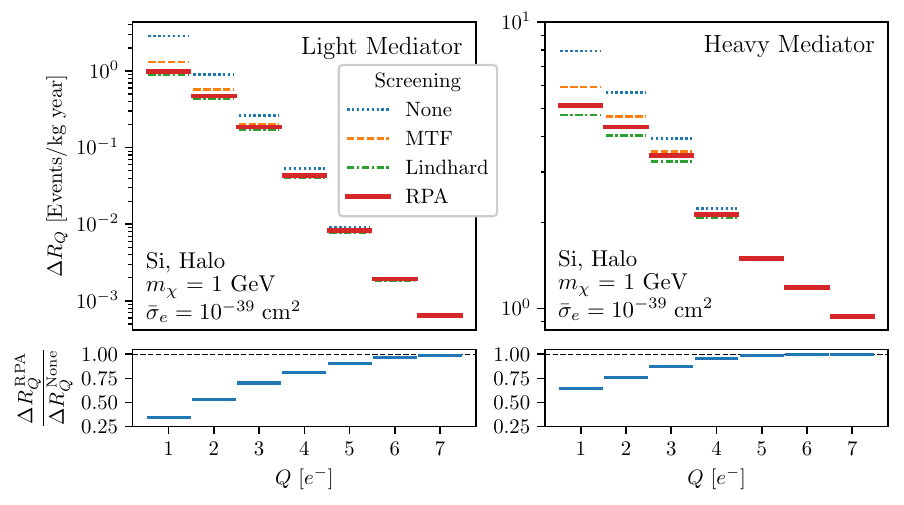}
        \caption{Electron recoil spectra in Si for halo DM with the screening models in Fig.~\ref{fig:S_screening}. We use the ionization model from~\cite{RK} at 100~K. The Lindhard model accurately replicates RPA screening for these examples. The bottom panel shows the ratio of the rates when including the RPA screening to not including any screening.} 
        \label{fig:recoil_spectrum_screening}
    \end{figure*}

    In Fig.~\ref{fig:recoil_spectrum_screening}, we compare the electron recoil spectra of Si with the four screening models introduced above for the $m_{\chi} = 1$ GeV halo DM example. We can see that the MTF model slightly overestimates the recoil rates, while the Lindhard model slightly underestimates them compared to RPA screening. Overall, the Lindhard model performs better than the MTF model, although the RPA dielectric function must be used in the numerator to fully capture the high momentum contributions.

    \begin{figure*}
        \includegraphics[width=1\linewidth]{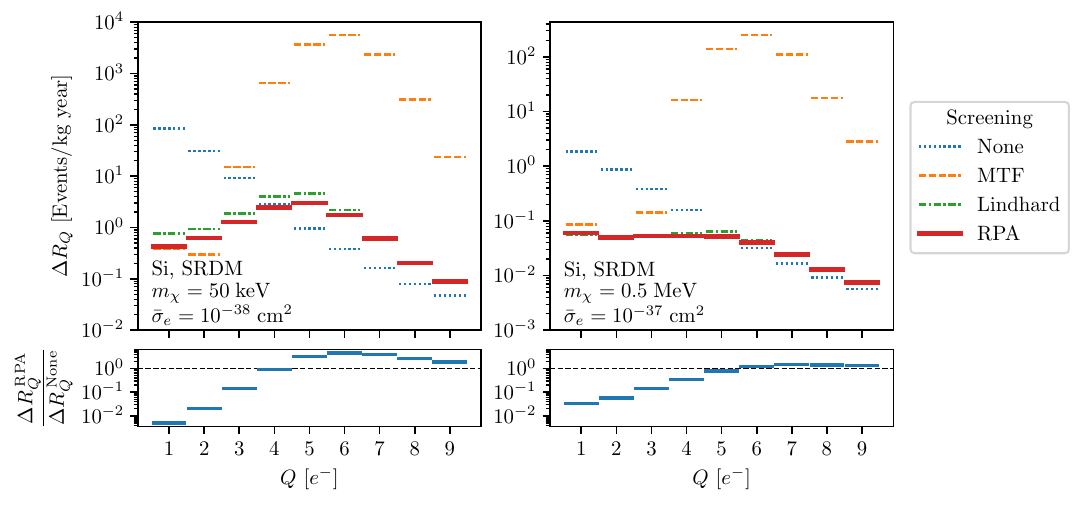}
        \caption{Electron recoil spectra in Si for two examples of solar-reflected DM fluxes, both assuming a massless mediator. As in Fig.~\ref{fig:recoil_spectrum_screening}, the Lindhard model comes closest to RPA screening; however, there are still inconsistencies in the $q \rightarrow 0$ limit, which contribute to the increased rate for low $Q$ in the left plot. The modified Thomas-Fermi (MTF) model should not be used for boosted DM (or other boosted dark-sector particles). The bottom panel shows the ratio of the rates when including the RPA screening to not including any screening.} 
        \label{fig:SRDM_recoil_spectrum_screening}
    \end{figure*}

    We show the different screening models applied to the two SRDM examples in Fig.~\ref{fig:SRDM_recoil_spectrum_screening}. The Lindhard model shows reasonable agreement with full RPA screening, since it replicates the plasmon peak. However, it performs slightly worse for $m_{\chi} = 50$~keV compared to $m_{\chi} = 0.5$~MeV, since the RPA dielectric function with LFEs predicts a lower plasmon peak than the Lindhard model in the extreme $q \rightarrow 0$ limit. This limit is included in the $m_{\chi} = 50$ keV phase space, but is excluded by the $m_{\chi} = 0.5$ MeV kinematic limits as can be seen in Fig.~\ref{fig:S_DM_limits}. The MTF model should not be used for boosted DM, since the plasmon region has divergent behavior; we include it here to demonstrate why the Lindhard model or RPA need to be used for correctly modeling high-velocity DM (or other boosted dark-sector particles).
    Finally, we include the spectrum with no screening to demonstrate that the implementation of some form of dielectric screening is even more important for boosted DM compared to halo DM, since it completely alters the shape of the recoil spectrum. The bottom plots show that the $Q = $ 1--2 bins are suppressed by two orders of magnitude when RPA screening is applied.

\subsection{Sensitivity projections} \label{sec:Results:Reach}
    In this subsection, we discuss the projected sensitivity in the $\bar{\sigma}_e$ versus DM mass $m_\chi$ plane for the different materials and DM fluxes considered above. Results from other codes discussed in the introduction will also be compared to our new \textsc{QCDark2} results for Si and Ge. We assume no backgrounds and a one-electron threshold such that the expected number of DM-electron scattering events can be calculated by simply integrating Eq.~(\ref{eq:general_rate}) over transferred energy $\omega$. The maximum energy transfer included for Si, Ge, GaAs, and SiC is 50 eV; for diamond we include up to 150 eV due to its much larger bandgap. All curves are calculated at a 90\% confidence level (C.L.), which corresponds to 2.3 events over an exposure of 1 kg year.

    \begin{figure*}
        \includegraphics[width=1\linewidth]{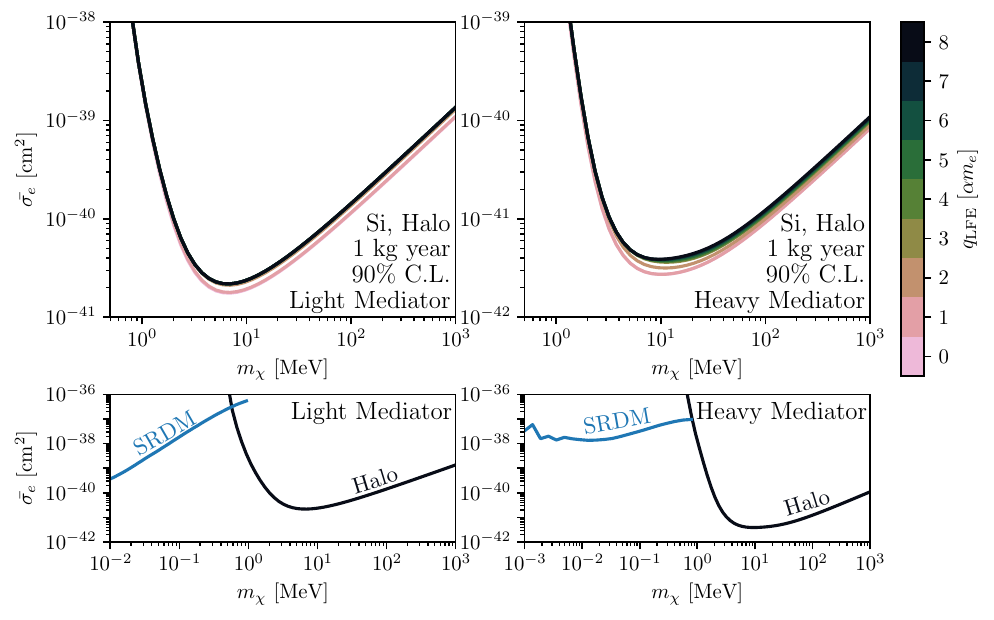}
        \caption{Projections for exclusion curves at 90\% C.L. in Si, assuming no backgrounds and a one-electron threshold. The top two plots show the reach for halo DM as LFEs are included up to varying $q_{\mathrm{LFE}}$. The bottom two plots show the projections for SRDM in the light and heavy mediator limits along with the full LFE screening halo projections.}
        \label{fig:projection_halo_lfe_cutoff_srdm_Si}
    \end{figure*}

    In Fig.~\ref{fig:projection_halo_lfe_cutoff_srdm_Si}, we show a detailed view of how the reach for Si changes as $q_{\mathrm{LFE}}$ is increased from 0 to 8 $\alpha m_e$ as was done in Fig.~\ref{fig:composite_epsilon_recoil_spectrum}. A 20\%--40\% reduction in the reach for halo DM occurs for masses larger than a few MeV for both the light and heavy mediator limits. This reduction is more gradual as a function of LFE cutoff for the heavy mediator case as we would expect from Fig.~\ref{fig:composite_epsilon_recoil_spectrum}. We observed that the SRDM limits are barely affected by the inclusion of LFEs ($<$ 5\% difference). This may seem unexpected since LFEs have such a dramatic effect on the plasmon peak in Fig.~\ref{fig:ELF_all_materials}, however they mainly redistribute the power away from the peak at low momentum instead of causing an overall reduction in the dynamic structure factor that we see at higher momentum. Experimental data currently show much larger backgrounds for the lowest $Q$ bins (especially $Q\le 2$) and moreover consider each $Q$ bin separately; in this case, the impact of using our updated calculations on the sensitivity is larger than shown by these plots.
    The SRDM projections shown here are calculated only for vector mediators, as the difference between vector and scalar mediator sensitivities differ by less than 0.5\% above $m_{\chi} = 10$ keV; and they agree closely with those calculated using \textsc{QCDark1} with Lindhard screening in~\cite{Emken2024}. We note that for the heavy-mediator case, DM masses lower than $\sim$5--10~MeV are generally disfavored from bounds on the number of relativistic degrees of freedom~\cite{Boehm:2013jpa}.

    \begin{figure*}
        \includegraphics[width=1\linewidth]{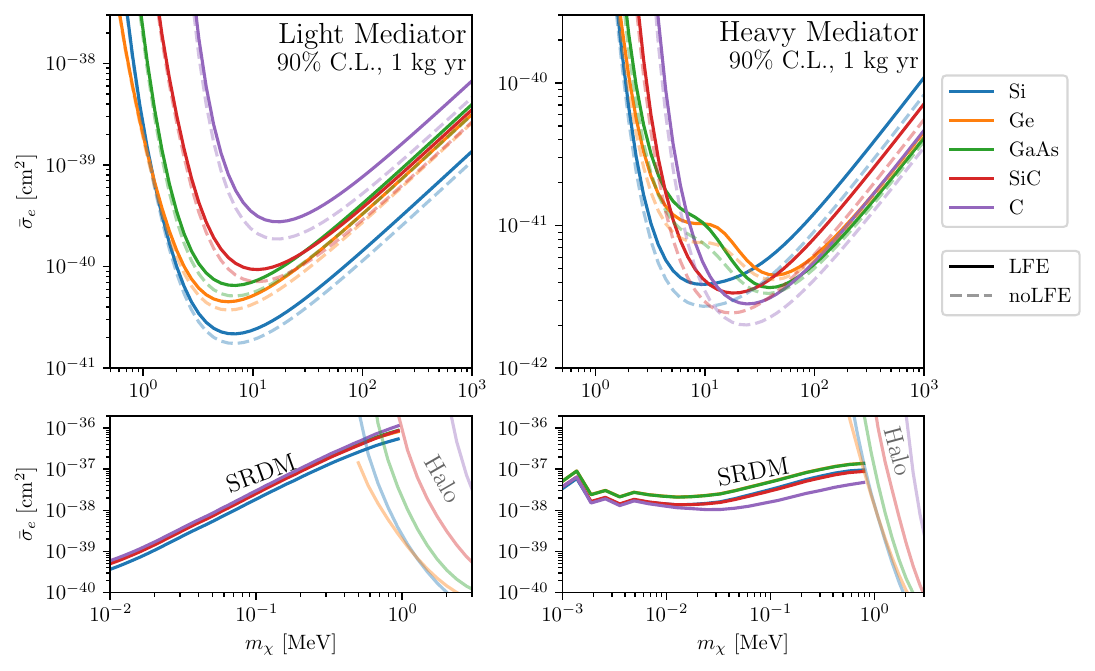}
        \caption{Projections for exclusion curves at 90\% C.L. in Si, Ge, GaAs, SiC, and diamond (C), assuming no backgrounds and a one-electron threshold. For the upper two plots, solid curves fully include LFEs (up to the maximum LFE cutoff for each material shown in Table~\ref{tab:parameters}) and the light dashed curves fully neglect LFEs. The lower left plot shows SRDM projections for all materials assuming a massless mediator, while the right plot shows the same for a heavy mediator. LFEs are fully included for all SRDM curves.}
        \label{fig:projection_all_materials}
    \end{figure*}

    Figure~\ref{fig:projection_all_materials} shows the projected reach for Si, Ge, GaAs, SiC, and diamond over an exposure of 1 kg year. Following the same behavior as Si in Fig.~\ref{fig:projection_halo_lfe_cutoff_srdm_Si}, LFEs reduce the reach by 20\%--50\% for DM masses larger than a few MeV for all materials. As for Si, there is again little change in the SRDM limits when the LFE cutoff is varied and so we only show the SRDM sensitivities with LFEs included. 
    
    Interestingly, the SRDM limits are very similar across materials (even if the rates are calculated per atom instead of per kg). We may be able to explain this similarity using a sum rule~\cite{CohenLouie} that relates the integrated loss function to the location of the plasmon peak $\omega_p$,
    \begin{equation} \label{eq:sum_rule}
        \int_0^\infty \omega \mathrm{Im}\left( \frac{-1}{\epsilon(\omega, q)} \right) d\omega = \frac{\pi}{2} \omega_p^2 \ .    
    \end{equation}
    Since all the materials we consider have similar plasmon peak shapes and locations, the integrated magnitudes of the loss functions are expected to be similar as well. Equation~(\ref{eq:sum_rule}) is technically only fulfilled for a complete basis set (and integration up to $\omega \rightarrow \infty$), but it applies well to our dielectric functions up to $q \lesssim 1\:\alpha m_e$, which covers the regions relevant to SRDM scattering. 

    \begin{figure*}
        \includegraphics[width=1\linewidth]{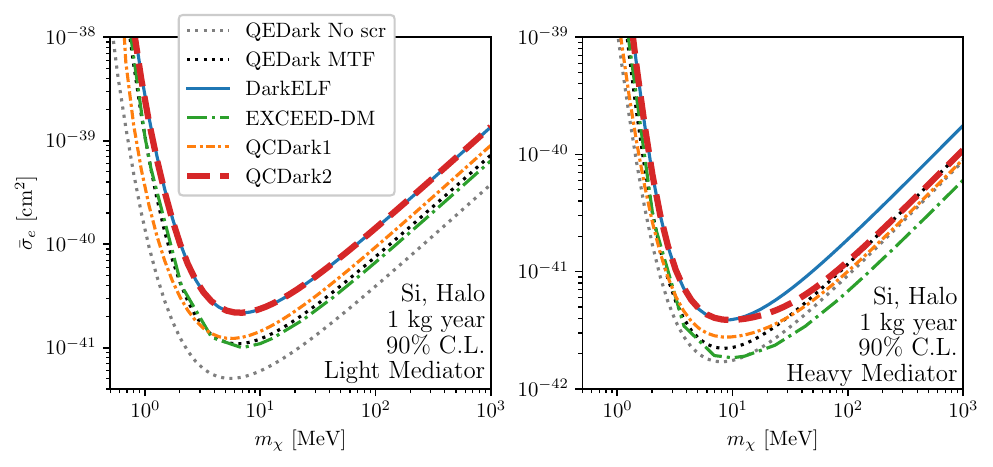}
        \caption{Projections for exclusion curves at 90\% C.L. in Si, assuming no backgrounds and a one-electron threshold. We compare all available codes including \textsc{QEDark} with no screening~\cite{QEDark} and with MTF screening, \textsc{DarkELF} with LFEs~\cite{DarkELF}, \textsc{EXCEED-DM} with its numerical dielectric function used for screening~\cite{EXCEED-DM_code}, \textsc{QCDark1} with MTF screening~\cite{QCDark1}, and \textsc{QCDark2} with LFEs (this work).}
        \label{fig:projection_all_codes_Si}
    \end{figure*}

\begin{figure*}
        \includegraphics[width=1\linewidth]{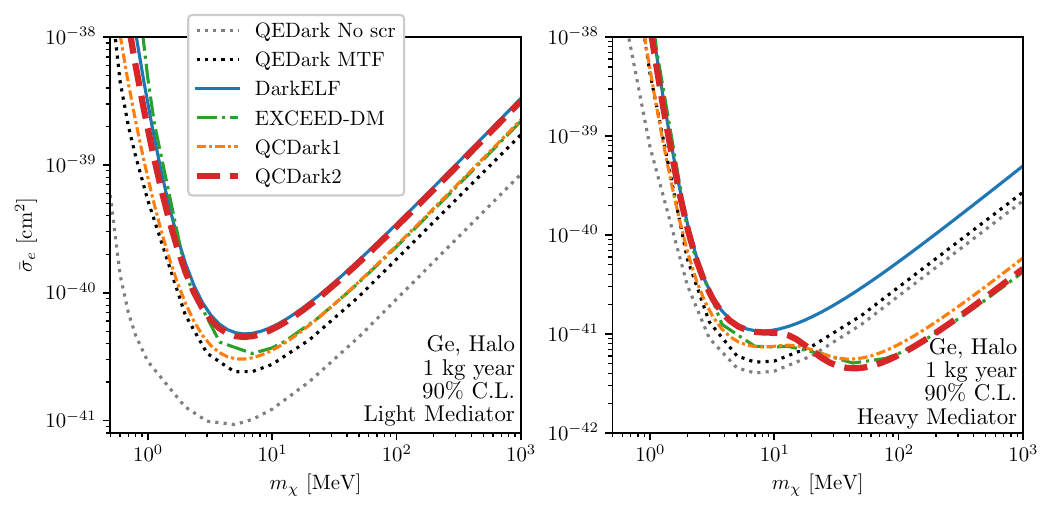}
        \caption{Projections for exclusion curves at 90\% C.L. in Ge, assuming no backgrounds and a one-electron threshold. We compare results from \textsc{QEDark} with no screening~\cite{QEDark} and with MTF screening, \textsc{DarkELF} with LFEs~\cite{DarkELF}, \textsc{EXCEED-DM} with its numerical dielectric function used for screening~\cite{EXCEED-DM_code}, \textsc{QCDark1} with MTF screening~\cite{QCDark1}, and \textsc{QCDark2} with LFEs (this work).}
        \label{fig:projection_all_codes_Ge}
    \end{figure*}

\begin{figure*}
        \includegraphics[width=1\linewidth]{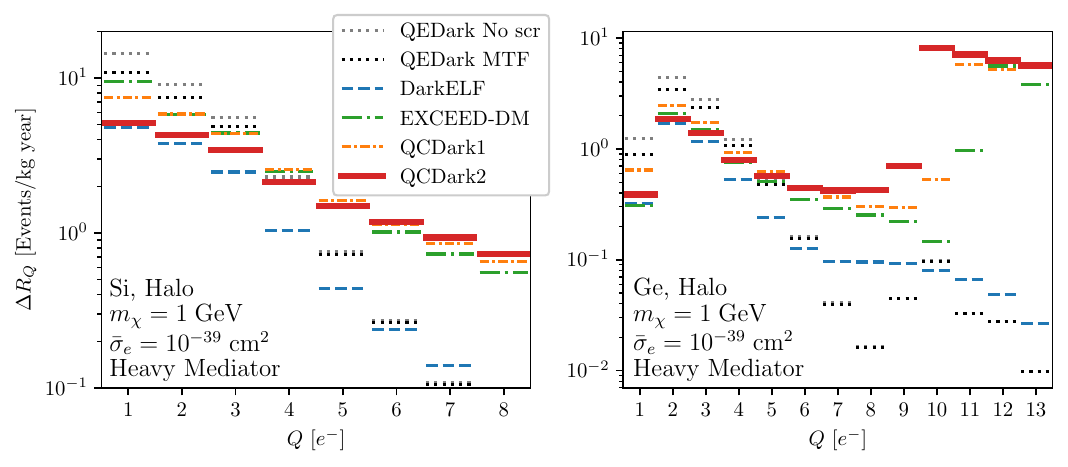}
        \caption{Electron recoil spectra in Si and Ge calculated with different DM-electron scattering codes for halo DM with a heavy mediator, using the ionization model from~\cite{RK} at 100 K for Si, and $Q(\omega) = 1 + \lfloor (\omega - \omega_{\mathrm{gap}}) / \omega_{pp} \rfloor$ for Ge, where $\omega_{\mathrm{gap}} = 0.67$~eV is the bandgap and $\omega_{pp} = 2.9$~eV is the pair creation energy. 
        We show \textsc{QEDark} with no screening~\cite{QEDark} and with MTF screening, \textsc{DarkELF} with LFEs~\cite{DarkELF}, \textsc{EXCEED-DM} with its numerical dielectric function used for screening~\cite{EXCEED-DM_code}, \textsc{QCDark1} with MTF screening~\cite{QCDark1}, and \textsc{QCDark2} with LFEs (this work).
        The high-$Q$ behavior of \textsc{QCDark1} and \textsc{EXCEED-DM} agree well with \textsc{QCDark2}, since these codes include high-momentum transitions and all-electron effects, while \textsc{DarkELF} is cut off at $q = 6\:\alpha m_e$ and \textsc{QEDark} does not include all-electron effects. The increased dielectric screening at low energy/momentum when LFEs are included is apparent from the reduction of the first $Q$ bin for \textsc{QCDark2} and \textsc{DarkELF}.}
        \label{fig:recoil_spectrum_codes}
    \end{figure*}

    In Figs.~\ref{fig:projection_all_codes_Si} and~\ref{fig:projection_all_codes_Ge}, we compare the halo DM sensitivities calculated using the other available DM-electron scattering codes for Si and Ge, respectively. Focusing first on Fig.~\ref{fig:projection_all_codes_Si}, we see that \textsc{QCDark2} and \textsc{DarkELF}\footnote{See Appendix \ref{sec:DarkELF_comparison} for a more detailed comparison of \textsc{QCDark2} and \textsc{DarkELF} with and without LFEs.} (both with LFEs) agree very well across the whole mass range in the light mediator limit. In the heavy mediator limit however, these start to diverge around 10~MeV. \textsc{DarkELF} predicts a shallower reach for these masses because their dielectric function is only tabulated up to $q = 6\:\alpha m_e$, while the \textsc{QCDark2} curve includes contributions up to $q = 25\:\alpha m_e$. The effect of neglecting these high-momentum contributions can also be seen in the electron recoil spectrum in the left plot of Fig.~\ref{fig:recoil_spectrum_codes}. Due to the kinematic limits on $(\omega, q)$ for halo DM, recoils with large transferred energy $\omega$ must necessarily transmit large momentum $q$ as well, so neglecting the high-momentum response significantly reduces the event rate for $Q \geq 3$.

    The lower reach of \textsc{QCDark1} is due to the MTF dielectric function used for screening and the neglect of LFEs; we showed in Fig.~\ref{fig:recoil_spectrum_screening} that MTF screening results in slightly larger scattering rates than RPA screening, and that LFEs reduce the scattering rate in Fig.~\ref{fig:composite_epsilon_recoil_spectrum}.
    The black \textsc{QEDark} curve also has lower reach due to MTF screening, and \textsc{QEDark} with no screening (as was initially calculated in \cite{QEDark}) in gray has even lower reach, showing the importance of including dielectric screening. There is a sharp turnover at 10~MeV for both \textsc{QEDark} curves because the loss function quickly vanishes above $q = 6\:\alpha m_e$ (see Fig.~5 in~\cite{QEDark}; the effect is even more obvious for Ge), since high-momentum modes are not included in its plane-wave basis sets. This results in a similar decrease in the $Q > 3$ bins of the electron recoil spectrum as we saw for \textsc{DarkELF} in Fig.~\ref{fig:recoil_spectrum_codes}. 
    The \textsc{EXCEED-DM} rates include extra valence to free electron transitions and are screened by a numerical dielectric function calculated with these extra states, also resulting in a lower reach.

    Moving on to the Ge results in Fig.~\ref{fig:projection_all_codes_Ge}, we see that the light mediator limit looks very similar to the various Si curves in Fig.~\ref{fig:projection_all_codes_Si}, and much of the same analysis applies here. There is a new feature in the projected reach for a heavy mediator model, where an enhancement of the reach to lower cross sections is seen for the \textsc{EXCEED-DM}, \textsc{QCDark1}, and \textsc{QCDark2} codes. This enhanced sensitivity is due to the onset of excitations from the $3d$ orbitals starting at $\omega = 26$ eV, which can be seen in the band structure shown in Appendix~\ref{sec:all_materials_details} (see Fig.~\ref{fig:bands_all_materials}). The $3d$ orbitals cause an increase in the rate of high-$Q$ events in the electron recoil spectrum shown in the right plot of Fig.~\ref{fig:recoil_spectrum_codes}. 
    The importance of the $3d$ orbitals for DM-electron scattering in Ge was first pointed out in~\cite{Lee:2015qva}, which used a semianalytic model to include these states.
    \textsc{QCDark1} and \textsc{QCDark2} both use an all-electron basis, which, by definition, includes the $3d$ orbitals as possible initial states $i$ in the calculation of the dielectric function in Eq.~(\ref{eq:RPA}). \textsc{EXCEED-DM} uses the projector augmented wave (PAW) method to split up the calculation of core and valence electronic states, using all-electron reconstruction to correctly model the $3d$ orbitals. \textsc{DarkELF} also uses DFT states calculated via the PAW method, however the $3d$ states are included only in the pseudopotential core and do not contribute to the scattering rate. \textsc{QEDark} does model the $3d$ orbitals separately from the pseudopotential, but the high-momentum modes that dominate the scattering rate are not captured due to the plane-wave basis set. This example of Ge particularly exemplifies how important the choice of DFT method is for capturing the relevant responses that contribute to the scattering rate.

\section{Discussion} \label{sec:Discussion}
    Before concluding, we would like to discuss some subtleties of the dielectric function and DFT calculations, recent work on generalized DM-electron interactions, the transverse dielectric response for ultrarelativistic DM, and the applicability of the \textsc{QCDark2} loss function to the neutrino fog and the Migdal effect. These topics serve to illustrate our work in the context of the wider fields of electronic structure calculations and DM-electron interactions, and offer some possible extensions based on recent ideas.

    \paragraph{DFT and the dielectric function.} 
    In this work, we have computed the dielectric function at the RPA level using all-electron Kohn-Sham wave functions, and have included LFEs up to fairly large momentum. In principle, one can apply improved theoretical methods to more accurately calculate the dielectric response. Improvements can include more intensive DFT calculations, such as using hybrid exchange-correlation functionals to accurately model the bandgap~\cite{QCDark1,EXCEED-DM_code} and applying a Hubbard $U$ correction to shift the $3d$ bands of Ge and GaAs~\cite{Catena2021}. More expensive improvements include energy corrections to the KS states through the $\mathrm{G}_0\mathrm{W}_0$ method~\cite{Peterson2023}, using time-dependent DFT to calculate the dielectric function beyond the RPA with contributions from the exchange-correlation kernel~\cite{Weissker2010}, and including excitonic effects in the dielectric function by solving the Bethe-Salpeter equation (BSE)~\cite{Taufertshofer2025}.

    The key challenge, however, is the accurate modeling of the electronic structure for the wide range of momenta and energies needed to calculate DM-electron scattering rates. 
    Indeed, we have found that the elements of the calculation required for accuracy over a wide range of materials and potential DM parameters (e.g., \textit{ab initio} treatments of the structure factor, high momentum cutoffs, LFEs, all-electron effects, dense $k$-grids) already carry a heavy computational burden, which makes using even more expensive electronic structure methods impractical (or even impossible). For this reason, it is crucial to explore and benchmark ad hoc corrections based on experimental data or higher-level calculations, as well as approximations based on the $(\omega, q)$ region of interest. For example, we use a scissor correction in this study to correct the bandgap, but tuned hybrid functionals or Hubbard $U$ corrections could be useful to obtain a more accurate initial electronic structure at a cost that still allows for accurate calculations of the scattering rate. In addition, we relied on concatenating the dielectric function with and without LFEs in different momentum regimes for computational efficiency. A similar procedure may be performed with \textit{ab initio} and model dielectric functions to further reduce the cost, or potentially RPA and GW+BSE dielectric functions to include excitonic effects in the $(\omega, q)$ regime where they have the largest effect. Finally, we believe that mitigating the cost of all-electron calculations with carefully benchmarked methods like PAW-based reconstructions~\cite{EXCEED-DM,EXCEED-DM_code} are a promising avenue.

    \paragraph{Generalized DM-electron interactions.} Generalized DM-electron interactions have been a topic of interest for many years~\cite{Catena2021,Krnjaic2025,Hochberg2025b}. While we have focused on a DM model that gives the usual heavy- and light-mediator form factors in Eq.~(\ref{eq:F_DM_limits}), other possible models exist with different couplings to the SM.\footnote{We note that a detailed study of UV completions of these generalized interactions may show that other experimental probes are more constraining than even optimistic projections for future direct-detection experiments.}
        
    It was pointed out in~\cite{Hochberg2025b} that any nonrelativistic DM-electron interactions that do not depend on the electron's spin (but may depend on the DM spin) can be described by the charge density response, or equivalently, by the dielectric function, which we have calculated in this work. The only models that are fully independent of the electron spin are scalar- and vector-mediated interactions, to which our analyses in this paper apply. However, partial interactions of anapole, electric dipole, and magnetic dipole couplings can also be calculated with the dielectric function, while the spin-density response is required to fully characterize these general DM-electron interactions. A different formalism to describe general interactions is developed in~\cite{Catena2021} and the evaluation of the generalized DM-electron interactions can be performed with an extension of \textsc{QEDark}, \textsc{QEDark-EFT}~\cite{QEDark-EFT}. This however still suffers from the same limitations of plane wave DFT as \textsc{QEDark}. 

    \paragraph{Transverse dielectric response.} 
    While the spin-dependent terms discussed above are not relevant to the simple vector or scalar mediator models considered in this study, an extra interaction term still must be taken into account when considering any model with highly relativistic DM. This term arises from the vector potential generated by a fast-moving DM particle, which induces a transverse response in the detector material in addition to the longitudinal Coulomb response originating from the scalar potential~\cite{Allison1980,Fano1963,Bichsel2020,DresselGruner}. At low speeds, the vector potential is negligible, and the dielectric response is completely captured by the loss function in Eq.~(\ref{eq:dynamic_structure_factor}). At ultrarelativistic speeds, we must modify the loss function to include the transverse response~\cite{Essig2024},
    \begin{multline} \label{eq:transverse_loss_function}
        \mathrm{Im} \left( \frac{-1}{\epsilon(\omega, q)} \right) \rightarrow 
        \mathrm{Im} \left( \frac{-1}{\epsilon_L(\omega, q)} \right) \\
        + (v^2 q^2 - \omega^2) \mathrm{Im} \left( \frac{1}{\epsilon_T(\omega, q) \omega^2 - q^2} \right) \ .
    \end{multline}
    The longitudinal dielectric function, $\epsilon_L$, is the same dielectric function we have calculated throughout this paper with RPA. While $\epsilon_T(\omega, \mathbf{q} \rightarrow 0) = \epsilon_L(\omega, \mathbf{q} \rightarrow 0)$, the transverse response quickly vanishes at larger $q$ due to the dispersion in the denominator. Additionally, at the velocities relevant for SRDM ($v \lesssim 0.2$), the integration limits of the cross section exclude any significant contributions from this term, so we neglect it throughout our analysis.\footnote{The transverse response was also shown to be many orders of magnitude smaller than the longitudinal response for bremsstrahlung produced by nuclear recoils in semiconductors using simple models for the loss functions in~\cite{Kozaczuk2020} (considering excitations caused by initial nuclear recoils allows one to probe the plasmon region at low momentum for halo DM, which is not kinematically allowed for the inelastic electronic collisions we consider in this paper).}
    In principle, the transverse dielectric function can be calculated at the RPA level~\cite{Sharma1981}, and this term can be included for extremely boosted DM.
    We should also note that Eq.~(\ref{eq:transverse_loss_function}) assumes that the detector material is \emph{isotropic}; for a crystalline material, there will be a few modifications. First, even in the highest-symmetry cubic crystals, where $\epsilon_L(\omega, \mathbf{q} \rightarrow 0)$ is isotropic, at finite $\mathbf{q}$ the longitudinal dielectric function is no longer isotropic. Second, for a crystal with lower symmetry, the dielectric function cannot be neatly decomposed into a longitudinal and transverse part and the response must instead be described by the full $3\times 3$ dielectric tensor~\cite{DresselGruner}.

    \paragraph{Neutrino fog and Migdal rates.} In addition to governing DM-electron scattering rates, the loss function is also used to calculate the neutrino fog and electronic excitation rates via the Migdal effect. The neutrino fog refers to the unavoidable background of solar neutrinos that scatter with nuclei and electrons in a detector. This background becomes important as we probe ever-smaller DM-electron cross sections~\cite{Essig2012, Essig2018, Wyenberg2018}. Coherent elastic neutrino-nucleus scattering (CE$\nu$NS) is the largest contribution to the neutrino fog. While the CE$\nu$NS rate itself does not depend on the loss function, the boundary of the neutrino fog can vary slightly depending on the calculation of the DM-electron scattering rate. A study of the CE$\nu$NS boundary for DM-electron scattering in Si found that the \textsc{DarkELF} and \textsc{EXCEED-DM} boundaries agree well within an order of magnitude~\cite{Carew2024}. Since the \textsc{QCDark2} projections in Fig.~\ref{fig:projection_all_codes_Si} agree very closely with \textsc{DarkELF} in the light mediator case and fall between the reach of \textsc{DarkELF} and \textsc{EXCEED-DM} at high DM masses in the heavy mediator case, we expect the same trends to carry over to a neutrino fog analysis. The background from neutrino-electron scattering does depend on the loss function and even accesses the plasmon peak region, since the neutrinos are relativistic, however it is still subdominant to the CE$\nu$NS background by many orders of magnitude as shown in~\cite{Essig2012,Essig2018} and analyzed in detail in~\cite{Dent2026}. We expect that neutrino-electron scattering rates calculated with \textsc{QCDark2} would agree well with the rates calculated in~\cite{Dent2026} with \textsc{DarkELF}, since both accurately model the plasmon region of the loss function.

    The loss function is also used to calculate the rate of the Migdal effect for DM models that couple to nuclei. The Migdal effect refers to the prompt transfer of energy from a recoiling nucleus to an atomic electron~\cite{Migdal1939, Ibe2018}. When these nuclear recoils occur in semiconductors, the subsequent excitation of the electron density is described by the loss function~\cite{Knapen2021, Liang2021, Berghaus:2022pbu}. 
    \textsc{DarkELF} can be used to calculate Migdal rates because the high-momentum response is suppressed (similar to the light mediator limit of DM-electron scattering). The Migdal rate is also suppressed by an additional factor of $\omega^{-4}$ compared to DM-electron scattering, so we expect even the Ge rates to agree very closely between \textsc{DarkELF} and \textsc{QCDark2}, since the contribution from the $3d$ orbitals becomes negligible.

\section{Conclusion} \label{sec:Conclusion}
    In this paper, we have studied the importance of accurate dielectric screening for different DM fluxes. We presented the impacts of local field effects on scattering rates, specifically in their application to electron recoil spectra, at both low momentum, where we observe a broadening of the plasmon peak, and at high momentum, where we see a suppression when LFEs are included. We demonstrated the sensitivity of halo DM to the high-momentum region and of different solar-reflected DM models to one (or, depending on the mass, both) of these regions by varying the LFE momentum cutoff. 
    We also compared the electron recoil spectra calculated with previously used dielectric screening models to those with full RPA screening. 
    Finally, we computed sensitivity projections for Si, Ge, SiC, GaAs, and diamond; showed the effects of the LFE momentum cutoff on the sensitivity in Si; and compared \textsc{QCDark2} to other DM-electron scattering codes, finding good agreement with \textsc{DarkELF} for low DM masses (where low-momentum screening is important) and \textsc{QCDark1} for larger DM masses (where high-momentum transitions are important), as well as qualitative agreement with \textsc{EXCEED-DM} electron recoil spectra.
    While we focused mainly on Si and Ge to compare with past results, we provide precomputed dielectric functions for Si, Ge, GaAs, SiC, and diamond as resources for future studies. \textsc{QCDark2} is freely available for others to use at \url{https://github.com/meganhott/QCDark2}, and we have provided guides for reproducing all results in this paper and for running calculations on new materials. 

\acknowledgments
    We thank Tanner Trickle for feedback about \textsc{EXCEED-DM} and for providing the projections in Figs.~\ref{fig:projection_all_codes_Si} and \ref{fig:projection_all_codes_Ge}, and Hailin Xu for useful discussions regarding the projections for solar-reflected dark matter. 
    C.E.D.\ acknowledges support from the National Science Foundation under Grant No.~DMR-2237674. The Flatiron Institute is a division of the Simons Foundation. R.E. and M.H. acknowledge support from DOE Grant DE-SC0025309 and Simons Investigator in Physics Awards~623940 and MPS-SIP-00010469. 
    M.V.F-S.\ acknowledges support from the National Science Foundation award DMR-2427902.
    We also thank Stony Brook Research Computing and Cyberinfrastructure, and the Institute for Advanced Computational Science at Stony Brook University for access to the high-performance SeaWulf computing system, which was made possible by a National Science Foundation grant No.~1531492.

\appendix

\section{Derivation of the scattering rate for relativistic DM} \label{sec:scattering_rate_derivation}
    Here we show in detail the derivation of Eq.~(\ref{eq:cross_section_3d}) and the simplifications that can be made for an isotropic detector material to obtain Eq.~(\ref{eq:cross_section_iso}).
    We can build the cross section for a general DM-electron interaction starting from the basic form for $2 \rightarrow 2$ scattering:
    \begin{equation} \label{eq:gen_cross_section_a}
    \begin{split}
        \sigma_{if} = \frac{1}{4 E_1 E_2 |\mathbf{v}_1 - \mathbf{v}_2|} 
        \int \frac{d^3 p_3}{(2\pi)^3 2 E_3} \frac{d^3 p_4}{(2\pi)^3 2 E_4} \\
        \times (2\pi)^4 \delta^4(p_3 + p_4 - p_1 - p_2) 
        |\mathcal{M}_{if}|^2 \ ,  
    \end{split}
    \end{equation}
    where this is for a specific initial state ($i$) to final state ($f$) electronic transition. The total cross section is obtained by summing all possible initial and final states. 
    The incoming and outgoing DM particle may be relativistic and has four-momenta
    \begin{align*}
        p_1 &= (E_{\chi}, \mathbf{p}\,) \quad (\mathrm{incoming \; DM})\\
        p_3 &= (E_{\chi}', \mathbf{p'}) \quad (\mathrm{outgoing \; DM})
    \end{align*}
    and the electron in the crystal is nonrelativistic with initial and final four-momenta\footnote{Since the electrons are in a periodic lattice, they do not have definite momenta, however the typical momentum should be around the Fermi momentum that is $\mathcal{O}(\alpha m_e) = \mathcal{O}(4\:\mathrm{keV})$, which is safely nonrelativistic. We will write expressions containing $\mathbf{k'}$ in terms of the transferred momentum $\mathbf{q} = \mathbf{k'} - \mathbf{k}$. The DM particle is also not directly interacting with a single electron, but rather with the electron density.}
    \begin{align*}
        p_2 &= (E_e, \mathbf{k}\,) \quad (\mathrm{initial \; e}^-) \\
        p_4 &= (E_e', \mathbf{k'}) \quad (\mathrm{final \; e}^-)
    \end{align*}
    Working in the rest frame of the detector, we can set $|\mathbf{v}_1 - \mathbf{v}_2| \approx v$, the velocity of the DM, and $E_e = E_e' = m_e$.
    We can multiply by $1 = \int d\omega \delta(\omega - (E_e' - E_e)) = \int d\omega \delta(\omega - (\omega_f - \omega_i))$ to obtain a differential cross section with respect to transferred energy $\omega$. To obtain the full differential cross section, we sum over all initial and final electron states:
    \begin{equation} \label{eq:cross_section_a}
    \begin{split}
        \frac{d\sigma}{d\omega} = \sum_{if} &\frac{1}{16 m_e^2 E_{\chi} v}
        \int \frac{d^3 p'}{(2\pi)^3} \frac{d^3 k'}{(2\pi)^3} \frac{1}{E_{\chi}'} \\
        &\times (2\pi)^4\delta(E_{\chi}' + E_e' - (E_{\chi} + E_e)) \\
        &\times \delta^3(\mathbf{p'} + \mathbf{k'} - (\mathbf{p} + \mathbf{k}\,)) \\
        &\times |\mathcal{M}_{if}|^2 \delta(\omega - (\omega_f - \omega_i)) \ .
    \end{split}
    \end{equation}
    
    We will first derive separately the amplitudes for models with vector/scalar mediators to show the origin of the $H_{A'}$ function defined in Eq.~(\ref{eq:H}).\footnote{Recall that $A'$ refers to a general mediator (either vector or scalar).}
    For fermionic DM particles scattering off an electron with initial state $\ket{i}$ and final state $\ket{f}$ though a vector mediator $V$, the amplitude $\mathcal{M}_{if}^{V}$ is
    \begin{equation} \label{eq:matrix_element_vector}
    \begin{split}
        \mathcal{M}_{if}^{V} = \Bar{u}_{\chi}^{s'}(\mathbf{p'}) & (-i g_{\chi} \gamma^{\mu}) u_{\chi}^s(\mathbf{p}\,) \\
        &\times \left( \frac{-i g_{\mu 0}}{\omega^2 - |\mathbf{q}|^2 - m_{V}^2} \right) \\
        &\times \left( -i2m_e g_e \mel{f}{n(\mathbf{q})}{i} \right) \ ,
    \end{split}
    \end{equation}
    where the $\gamma^0$ term is dominant for the nonrelativistic electron vertex and we have taken the nonrelativistic limit of the electron spinors:
    \begin{equation*}
    \begin{split}
        \Bar{u}_e^{s'}(\mathbf{k}') (-i g_e \gamma^0) u_e^s(\mathbf{k}\,) \rightarrow 
        &-i 2 m_e g_e \mel{f}{n(\mathbf{k} - \mathbf{k'})}{i} \\
        &= -i 2 m_e g_e \mel{f}{n(\mathbf{q})}{i} \ .
    \end{split}
    \end{equation*}
    The factor of $2m_e$ comes from the normalization of the relativistic wave functions. Since the electrons are in a crystal, we have to replace the usual delta function over incoming and outgoing spins with an interaction with the electron density operator, $n(\mathbf{q}) = \sum_{\mathbf{r}_j} e^{-i \mathbf{q}\cdot\mathbf{r}_j}$.

    The spin-averaged amplitude is
    \begin{equation*}
    \begin{split}
        |\mathcal{M}_{if}^{V}|^2 &= (g_e g_{\chi})^2
        \left( \frac{1}{\omega^2 - |\mathbf{q}|^2 - m_{V}^2} \right)^2 \\
        &\quad \times \frac{1}{2} \mathrm{Tr} \sum_{s,s'} 
        \Bar{u}_{\chi}^{s'}(\mathbf{p'}) \gamma^0 u_{\chi}^s(\mathbf{p}\,) 
        \Bar{u}_{\chi}^{s}(\mathbf{p}\,) \gamma^0 u_{\chi}^{s'}(\mathbf{p'}) \\
        &\quad \times 4 m_e^2 \mel{f}{n(\mathbf{q})}{i}\mel{i}{n(-\mathbf{q})}{f} \ ,
    \end{split}
    \end{equation*}
    where we can simplify the spin sum 
    \begin{multline*}
        \frac{1}{2} \mathrm{Tr} \sum_{s,s'} 
        \Bar{u}_{\chi}^{s'}(\mathbf{p'}) \gamma^0 u_{\chi}^s(\mathbf{p}\,) 
        \Bar{u}_{\chi}^{s}(\mathbf{p}\,) \gamma^0 u_{\chi}^{s'}(\mathbf{p'}) \\
        = (E_{\chi} + E_{\chi}')^2 - |\mathbf{q}|^2 \ ,
    \end{multline*}
    resulting in
    \begin{multline} \label{eq:amplitude_fermion}
        |\mathcal{M}_{if}^{V}|^2 = 4(g_e g_{\chi})^2
        \frac{(E_{\chi} + E_{\chi}')^2 - |\mathbf{q}|^2}{(\omega^2 - |\mathbf{q}|^2 - m_{V}^2)^2} \\
        \times m_e^2 \mel{f}{n(\mathbf{q})}{i}\mel{i}{n(-\mathbf{q})}{f} \ .    
    \end{multline}

    A similar procedure can be done to obtain the amplitude when the dark mediator is a scalar, $\phi$. In this case, the amplitude is
    \begin{multline}
        \mathcal{M}_{if}^{\phi} = \Bar{u}_{\chi}^{s'}(\mathbf{p'}) (-i g_{\chi}) u_{\chi}^s(\mathbf{p}\,)
        \left( \frac{i}{\omega^2 - |\mathbf{q}|^2 - m_{\phi}^2} \right) \\
        \times \left( -i2m_e g_e \mel{f}{n(\mathbf{q})}{i} \right) \ ,    
    \end{multline}
    the spin-averaged amplitude is 
    \begin{equation*}
    \begin{split}
        |\mathcal{M}_{if}^{\phi}|^2 &= (g_e g_{\chi})^2
        \left( \frac{1}{\omega^2 - |\mathbf{q}|^2 - m_{\phi}^2} \right)^2 \\
        &\quad \times \frac{1}{2} \mathrm{Tr} \sum_{s,s'} 
        \Bar{u}_{\chi}^{s'}(\mathbf{p'}) u_{\chi}^s(\mathbf{p}\,) 
        \Bar{u}_{\chi}^{s}(\mathbf{p}\,) u_{\chi}^{s'}(\mathbf{p'}) \\
        &\quad \times 4 m_e^2 \mel{f}{n(\mathbf{q})}{i}\mel{i}{n(-\mathbf{q})}{f} \ ,
    \end{split}
    \end{equation*}
    and the spin sum simplifies to 
    \begin{multline*}
        \frac{1}{2} \mathrm{Tr} \sum_{s,s'} 
        \Bar{u}_{\chi}^{s'}(\mathbf{p'}) u_{\chi}^s(\mathbf{p}\,) 
        \Bar{u}_{\chi}^{s}(\mathbf{p}\,) u_{\chi}^{s'}(\mathbf{p'}) \\
        = 4 m_{\chi}^2 - (E_{\chi} - E_{\chi}')^2 + |\mathbf{q}|^2 \ .
    \end{multline*}
    Therefore, the spin-averaged amplitude for a scalar mediator is 
    \begin{multline} \label{eq:amplitude_scalar}
        |\mathcal{M}_{if}^{\phi}|^2 = 4(g_e g_{\chi})^2
        \frac{4 m_{\chi}^2 - (E_{\chi} - E_{\chi}')^2 + |\mathbf{q}|^2}{(\omega^2 - |\mathbf{q}|^2 - m_{\phi}^2)^2} \\
        \times m_e^2 \mel{f}{n(\mathbf{q})}{i}\mel{i}{n(-\mathbf{q})}{f} \ ,    
    \end{multline}
    and it is clear that $H_{A'}$ is simply the spin sum for DM-electron interactions (within a crystal) in each model. 

    We can write a unified amplitude in terms of $H_{A'}(\mathbf{q})$ and plug this back into Eq.~(\ref{eq:cross_section_a}),
    \begin{equation*}
    \begin{split}
        \frac{d\sigma}{d\omega} &= \frac{(g_e g_{\chi})^2}{4 E_{\chi} v} 
        \int \frac{d^3 p'}{(2\pi)^3} \frac{d^3 k'}{(2\pi)^3} \frac{1}{E_{\chi}'} \\
        &\quad \times (2\pi)^4 \delta(E_{\chi}' + E_e' - E_{\chi} - E_e) \\
        &\quad \times \delta^3(\mathbf{p}' + \mathbf{k}' - \mathbf{p} - \mathbf{k})
        \frac{H_{A'}(\mathbf{q})}{(\omega^2 - |\mathbf{q}|^2 - m_{A'}^2)^2} \\
        &\quad \times \sum_{if} \mel{f}{n(\mathbf{q})}{i}\mel{i}{n(-\mathbf{q})}{f} \delta(\omega - (\omega_f - \omega_i)) \ ,
    \end{split}
    \end{equation*}
    then use momentum conservation to integrate over $\mathbf{k}'$ and substitute $d^3p' \rightarrow d^3q$:
    \begin{equation*}
    \begin{split}
        \frac{d\sigma}{d\omega} &= \frac{2\pi(g_e g_{\chi})^2}{4 E_{\chi} v} 
        \int \frac{d^3 q}{(2\pi)^3} \frac{1}{E_{\chi}'} \\
        &\quad \times \delta(E_{\chi}' + E_e' - E_{\chi} - E_e)
        \frac{H_{A'}(\mathbf{q})}{(\omega^2 - |\mathbf{q}|^2 - m_{A'}^2)^2} \\
        &\quad \times \sum_{if} \mel{f}{n(\mathbf{q})}{i}\mel{i}{n(-\mathbf{q})}{f} \delta(\omega - (\omega_f - \omega_i)) \ .
    \end{split}
    \end{equation*}
    The dynamic structure factor is defined as in~\cite{DarkELF}:
    \begin{equation*}
        S(\omega, \mathbf{q}) = \frac{2\pi}{V_T} \sum_{if} |\mel{f}{n(-\mathbf{q})}{i}|^2 \delta(\omega - (\omega_f - \omega_i))\ ,     
    \end{equation*}
    where $V_T$ is the total volume of the crystal. The rate is typically reported as a function of the total mass rather than the total volume, so we can replace $V_T$ with $m_T/\rho_T$, where $m_T$ is the total crystal mass and $\rho_T$ is the mass density. Our cross section is now
    \begin{multline} \label{eq:cross_section_S_a}
        \frac{d\sigma}{d\omega} = 
        \frac{m_T}{\rho_T} \frac{(g_e g_{\chi})^2}{4 E_{\chi} v} 
        \int \frac{d^3 q}{(2\pi)^3}\frac{1}{E_{\chi}'}
        \delta(E_{\chi}' + E_e' - E_{\chi} - E_e) \\
        \times \frac{H_{A'}(\mathbf{q})}{(\omega^2 - |\mathbf{q}|^2 - m_{A'}^2)^2} S(\omega, \mathbf{q}) \ .
    \end{multline}
    Soft collisions dominate between a relativistic scatterer and the electrons in a crystal~\cite{Allison1980}, so we can make the assumption that $\omega \ll (E_{\chi} = \gamma m_{\chi})$ and $q \ll (p = \gamma m_{\chi} v)$. We can apply this to the remaining delta function and write its argument in terms of the angle $\theta$ between $\mathbf{v}$ and $\mathbf{q}$:
    \begin{align} \label{eq:delta_angle}
        \delta(E_{\chi}'& + E_e' - E_{\chi} - E_e) \nonumber \\
        &= \delta\left(\sqrt{m_{\chi}^2 + |\mathbf{p} - \mathbf{q}|^2|} - \sqrt{m_{\chi}^2 + |\mathbf{p}\,|^2|} + \omega\right) \nonumber \\
        &= \begin{multlined}[t] \delta\left(\sqrt{m_{\chi}^2 + |\mathbf{p}\,|^2 - 2\gamma m_{\chi} v q \cos{\theta} + |\mathbf{q}|^2} \right. \nonumber \\
        \left. - \sqrt{m_{\chi}^2 + |\mathbf{p}\,|^2} + \omega \right) \nonumber \end{multlined} \\
        &= \delta\left( \frac{-2\gamma m_{\chi} v q \cos{\theta} + q^2}{2\sqrt{m_{\chi}^2 + p^2}} + \omega\right) \nonumber \\ 
        &= \delta\left(-v q \cos{\theta} + \frac{q^2}{2\gamma m_{\chi}} + \omega\right) \nonumber \\ 
        &= \frac{1}{qv}\delta\left(\cos{\theta} - \frac{q}{2\gamma m_{\chi} v} - \frac{\omega}{qv}\right) \ .
    \end{align}
    This looks the same as the nonrelativistic derivation~\cite{QEDark,Trickle2020} up to the factor of $q/ 2\gamma m_{\chi} v$ that replaces $q/ 2m_{\chi} v$. If we make the approximation that the dynamic structure factor is isotropic [see the discussion around Eq.~(\ref{eq:dynamic_structure_factor})], we can integrate over the $\mathbf{q}$ angles to obtain
    \begin{multline} \label{eq:cross_section_v_a}
        \frac{d\sigma}{d\omega} = 
        \frac{m_T}{\rho_T} \frac{(g_e g_{\chi})^2}{16 \pi^2  E_{\chi}}
        \int_{q_{\mathrm{min}}}^{q_{\mathrm{max}}} dq\frac{q}{(E_{\chi} - \omega)} S(\omega, q) \\
        \times \frac{H_{A'}(q)}{(\omega^2 - q^2 - m_{A'}^2)^2}
        \frac{\Theta(v - v_{\mathrm{min}})}{v^2} \ ,
    \end{multline}
    where
    \begin{equation*}
        v_{\mathrm{min}} = \frac{q}{2\gamma m_{\chi}} + \frac{\omega}{q} \ .
    \end{equation*}
    We can also use energy conservation at the maximum momentum transfer to determine the integration bounds on $q$ in Eq.~(\ref{eq:cross_section_v_a}):
    \begin{align} \label{eq:q_lim_a}
    \begin{split}
        q_{\mathrm{min}} &= \gamma m_{\chi} v - \sqrt{(\gamma m_{\chi} - \omega)^2 - m_{\chi}^2} \ , \\
        q_{\mathrm{max}} &= \gamma m_{\chi} v + \sqrt{(\gamma m_{\chi} - \omega)^2 - m_{\chi}^2} \ .
    \end{split}
    \end{align}
    
    This cross section can be plugged into Eq.~(\ref{eq:general_rate}) and integrated over $v$ with the flux if the energy of the DM particle is written in terms of its velocity, $E_{\chi} = \gamma m_{\chi} = m_{\chi}/\sqrt{1 - v^2}$. The step function obtained by integrating over the angular variables becomes redundant with the explicitly calculated integration bounds, so we will remove it from the cross section for now. The standard scattering rate derivation for halo DM uses the step function to simplify the velocity integration, so we show it in Eq.~(\ref{eq:cross_section_v_a}) for clarity.

    Writing in terms of the standard reference cross section for a general dark mediator mass~\cite{QEDark},
    \begin{equation*}
        \Bar{\sigma}_e = \frac{\mu_{\chi e}^2 (g_e g_{\chi})^2}{\pi (m_{A'}^2 + (\alpha m_e)^2)^2} \ ,
    \end{equation*}
    and writing the dynamic structure factor in terms of the loss function using Eq.~(\ref{eq:dynamic_structure_factor}), the cross section can be represented as 
    \begin{multline} \label{eq:cross_section_sigma_a}
        \frac{d\sigma}{d\omega} = 
        \frac{m_T}{\rho_T} \frac{\Bar{\sigma}_e}{32 \pi^2 \alpha \mu_{\chi e}^2 E_{\chi} v^2}
        \int_{q_{\mathrm{min}}}^{q_{\mathrm{max}}} dq \:q^3 \frac{H_{A'}(q)}{(E_{\chi} - \omega)} \\
        \times \frac{(m_{A'}^2 + (\alpha m_e)^2)^2}{(\omega^2 - |\mathbf{q}|^2 - m_{A'}^2)^2} 
        \mathrm{Im} \left( \frac{-1}{\epsilon(\omega, q)} \right) \ .
    \end{multline}

    As discussed in Section~\ref{sec:Discussion}, the transverse dielectric response must be included in addition to the longitudinal response for ultrarelativistic DM, which amounts to modifying the loss function with an additional term [see Eq.~(\ref{eq:transverse_loss_function})].
    Considering the opposite limit, if we assume the incoming DM particles are nonrelativistic as is the case for halo DM, we obtain the usual expression for the effective field theory description of fermionic DM-electron interactions~\cite{Trickle2020}:
    \begin{align*}
        \frac{d\sigma}{d\omega} &= \begin{multlined}[t]
        \frac{m_T}{\rho_T} \frac{\Bar{\sigma}_e}{32 \pi^2 \alpha \mu_{\chi e}^2 v^2}
        \int_{q_{\mathrm{min}}}^{q_{\mathrm{max}}} dq \:q^3 \frac{4 m_{\chi}^2}{m_{\chi}^2}
        \frac{(m_{A'}^2 + (\alpha m_e)^2)^2}{(q^2 + m_{A'}^2)^2} \\
        \times \mathrm{Im} \left( \frac{-1}{\epsilon(\omega, q)} \right) 
        \end{multlined} \\
        &= \frac{m_T}{\rho_T} \frac{\Bar{\sigma}_e}{8 \pi^2 \alpha \mu_{\chi e}^2 v^2}
        \int_{q_{\mathrm{min}}}^{q_{\mathrm{max}}} dq \:q^3 |F_{\mathrm{DM}}(q)|^2\:
        \mathrm{Im} \left( \frac{-1}{\epsilon(\omega, q)} \right)
    \end{align*}

\section{Band structures and dynamic structure factors} \label{sec:all_materials_details}
\begin{figure*}[ht]
        \includegraphics[width=\linewidth]{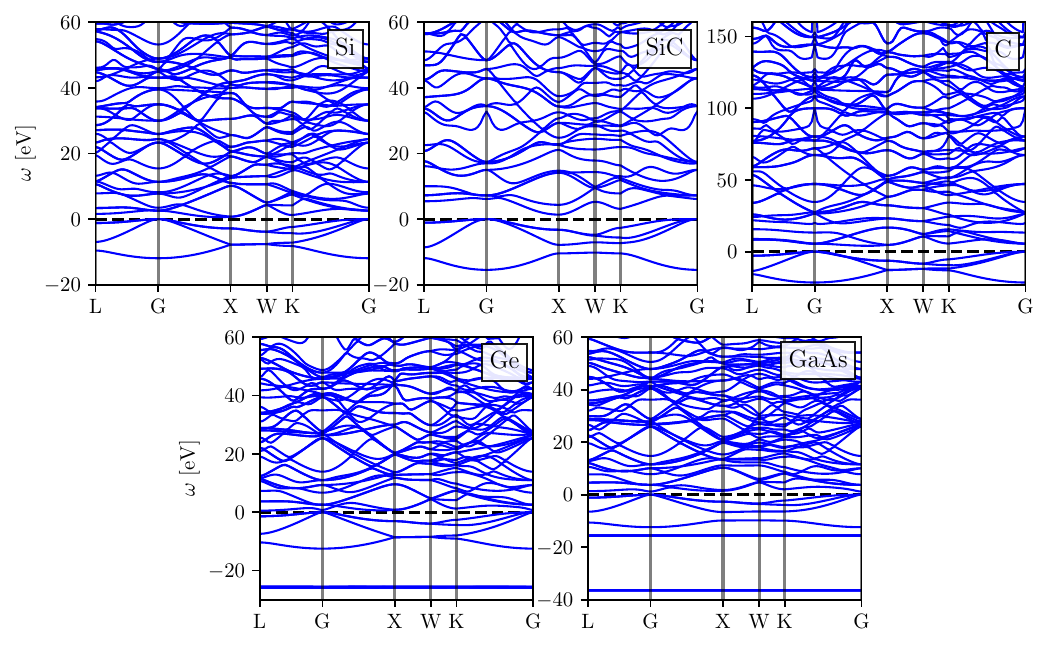}
        \caption{Band structures for all materials studied, where the dashed lines at $\omega = 0$ eV represent the top of the valence band. Conduction bands have been shifted according to the experimental band gaps listed in Table~\ref{tab:parameters}. The $3d$ orbitals of Ge and GaAs are located at $-25.8$~eV and $-15.5$~eV respectively. GaAs has further semicore states at $-36.4$~eV that also contribute to the dielectric function at high energy transfers.}
        \label{fig:bands_all_materials}
    \end{figure*}

\begin{figure*}[ht]
        \includegraphics[width=\linewidth]{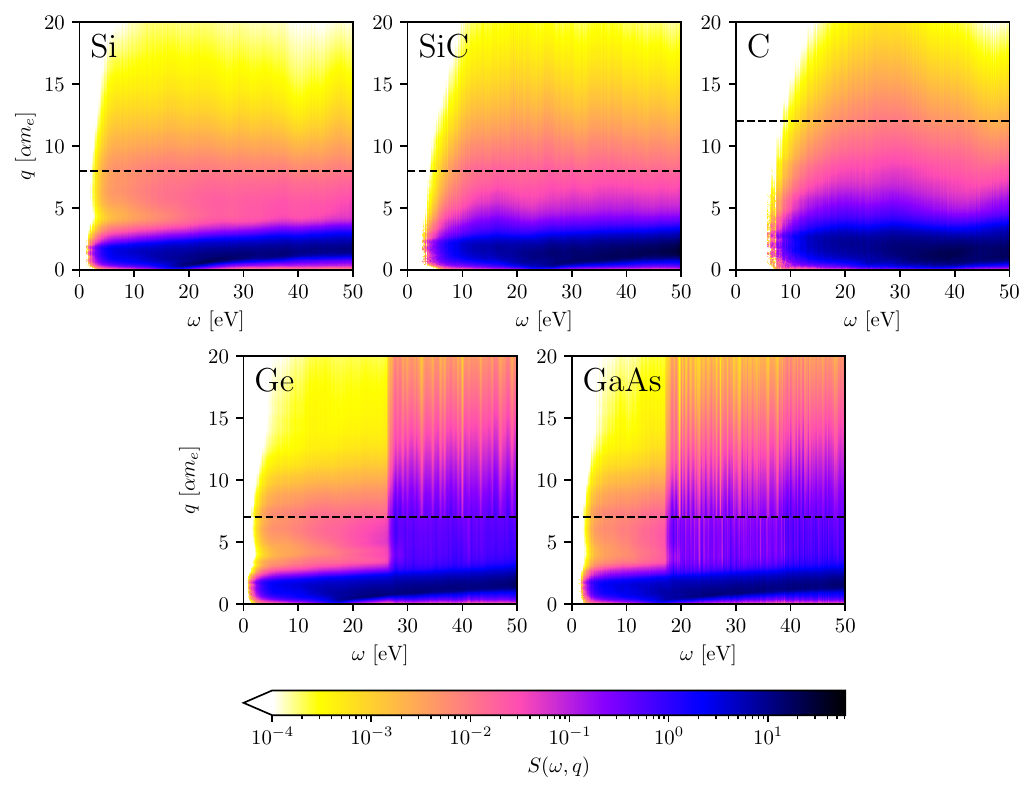}
        \caption{The dynamic structure factors for all materials studied, with dashed lines showing the LFE $q$ cutoff. The increased magnitude around 15--30 eV for Ge and GaAs are due to transitions from the $3d$ orbitals; these states can be seen in the band structure in Fig.~\ref{fig:bands_all_materials}.}
        \label{fig:S_all_materials}
    \end{figure*}

    We show the band structures for Si, Ge, GaAs, SiC, and diamond (C) in Fig.~\ref{fig:bands_all_materials} and their dynamic structure factors in Fig.~\ref{fig:S_all_materials}.
    
\section{Optical limit of the RPA dielectric function} \label{sec:optical_limit}
    Care must be taken when calculating $\epsilon_{00}$, $\epsilon_{0\mathbf{G'}}$, and $\epsilon_{\mathbf{G}0}$ in the $\mathbf{q} \rightarrow 0$ limit (also called the optical limit) due to the $|\mathbf{q}|$ factors in the denominator of Eq.~(\ref{eq:RPA}). We can cancel out this divergence by expanding the overlap terms with respect to a small perturbation in $\mathbf{q}$~\cite{Baroni1986,Hybertsen1987,Gajdos2006}. Since we are working in a periodic crystal, the molecular orbitals $\psi_{i\mathbf{k}}$ can be written in Bloch form, 
    \begin{equation*}
        \psi_{i\mathbf{k}}(\mathbf{r}\,) = u_{i\mathbf{k}}(\mathbf{r}\,) e^{i\mathbf{k}\cdot\mathbf{r}} \ ,    
    \end{equation*}
    where $u_{i\mathbf{k}}(\mathbf{r}\,)$ are orthonormal and periodic with the lattice. Now we can write the overlap integrals in terms of the Bloch functions
    \begin{multline*}
        \mel{i\mathbf{k}}{e^{-i\mathbf{q}\cdot\mathbf{r}}}{j(\mathbf{k}+\mathbf{q})} \\
        \begin{aligned}
        &= \mel{u_{i\mathbf{k}}}{e^{-i\mathbf{k}\cdot\mathbf{r}} e^{-i\mathbf{q}\cdot\mathbf{r}} e^{i(\mathbf{k}+\mathbf{q})\cdot\mathbf{r}}}{u_{j(\mathbf{k}+\mathbf{q})}} \\
        &= \braket{u_{i\mathbf{k}}}{u_{j(\mathbf{k}+\mathbf{q})}} \ ,
        \end{aligned} 
    \end{multline*}
    and expand $\ket{u_{j(\mathbf{k}+\mathbf{q})}}$ for small $\mathbf{q}$, 
    \begin{equation*}
        \ket{u_{j(\mathbf{k}+\mathbf{q})}} = \ket{u_{j\mathbf{k}}} + \sum_{m \neq j} \frac{\mel{u_{m\mathbf{k}}}{\Tilde{V}}{u_{j\mathbf{k}}}}{\omega_{j\mathbf{k}} - \omega_{m\mathbf{k}}} \ket{u_{m\mathbf{k}}} \ ,
    \end{equation*}
    where $\Tilde{V}$ is the perturbing potential for a system with Hamiltonian $\mathcal{H}$, 
    \begin{equation} \label{eq:V_perturb}
        \Tilde{V} = \mathcal{H}(\mathbf{k}+\mathbf{q}) - \mathcal{H}(\mathbf{k}\,) \ .
    \end{equation}
    The Bloch functions are orthonormal, so the perturbed overlap is
    \begin{equation} \label{eq:ol_overlap}
        \braket{u_{i\mathbf{k}}}{u_{j(\mathbf{k}+\mathbf{q})}} = 
        \frac{\mel{u_{i\mathbf{k}}}{\Tilde{V}}{u_{j\mathbf{k}}}}{\omega_{j\mathbf{k}} - \omega_{i\mathbf{k}}} \ .
    \end{equation}
    The molecular orbitals $\psi_{i\mathbf{k}}$ are eigenstates of $\mathcal{H} = \nabla^2 / 2m + V(\mathbf{r}\,)$, so we can apply the Hamiltonian and use the Bloch expansion to obtain
    \begin{equation*}
        \mathcal{H}(\mathbf{k}\,) u_{i\mathbf{k}}(\mathbf{r}\,) = \left[ -\frac{1}{2m_e} (\nabla + i\mathbf{k}\,)^2 + V(\mathbf{r}\,) \right] u_{i\mathbf{k}}(\mathbf{r}\,) \ .
    \end{equation*}
    Now the perturbation in Eq.~(\ref{eq:V_perturb}) can be written as 
    \begin{align*}
        \Tilde{V} &= \begin{multlined}[t]
        -\frac{1}{2m_e} (\nabla + i(\mathbf{k}+\mathbf{q}))^2 + V(\mathbf{r}\,) \\ 
        - \left( -\frac{1}{2m_e} (\nabla + i\mathbf{k}\,)^2 + V(\mathbf{r}\,) \right)
        \end{multlined} \\
        &= -\frac{i}{m_e} \mathbf{q} \cdot (\nabla + i\mathbf{k}\,) \ , 
    \end{align*}
    after dropping the $\mathcal{O}(q^2)$ term.
    Plugging $\Tilde{V}$ into Eq.~(\ref{eq:ol_overlap}) and converting back to molecular orbitals, the overlaps in the optical limit are
    \begin{equation}
        \mel{i\mathbf{k}}{e^{-i\mathbf{q}\cdot\mathbf{r}}}{j(\mathbf{k} + \mathbf{q})} \rightarrow -\frac{i}{m_e} \frac{\mathbf{q} \cdot \mel{i\mathbf{k}}{\nabla}{j\mathbf{k}}}{\omega_{j\mathbf{k}} - \omega_{i\mathbf{k}}} \ .
    \end{equation}
    The magnitude of the $\mathbf{q}$ factor cancels the diverging factor of $|\mathbf{q}|$ in the denominator of Eq.~(\ref{eq:RPA}). The overlaps $\mel{i\mathbf{k}}{\nabla}{j\mathbf{k}}$ are well defined (and easily calculated with existing \textsc{PySCF} routines).
    The head ($\epsilon_{00}$) and wings ($\epsilon_{0\mathbf{G'}}$ and $\epsilon_{\mathbf{G}0}$) of the RPA dielectric function in the proper optical limit are
    \begin{widetext}
    \begin{align}
        \epsilon_{00}(\omega, \mathbf{q} \rightarrow 0) &=    
        1 + \frac{4\pi \alpha}{V_{\mathrm{cell}} m_e^2} 
        \sum_{ij\mathbf{k}} \frac{f_{i\mathbf{k}} - f_{j\mathbf{k}}}{(\omega_{j\mathbf{k}} - \omega_{i\mathbf{k}})^2} 
        \lim_{\eta \rightarrow 0} 
        \frac{\left| \Hat{q} \cdot \mel{i\mathbf{k}}{\nabla}{j\mathbf{k}} \right|^2}{\omega - (\omega_{j\mathbf{k}} - \omega_{i\mathbf{k}}) + i \mathrm{sgn}(\omega_{j\mathbf{k}} - \omega_{i\mathbf{k}}) \eta} \\
        \epsilon_{0\mathbf{G'}}(\omega, \mathbf{q} \rightarrow 0) &=    
        \frac{4\pi \alpha \: i}{V_\mathrm{cell} m_e |\mathbf{G'}|} 
        \sum_{ij\mathbf{k}} \frac{f_{i\mathbf{k}} - f_{j\mathbf{k}}}{\omega_{j\mathbf{k}} - \omega_{i\mathbf{k}}}
        \lim_{\eta \rightarrow 0} 
        \frac{\Hat{q} \cdot \mel{i\mathbf{k}}{\nabla}{j\mathbf{k}} \mel{j\mathbf{k}}{e^{i \mathbf{G'}\cdot\mathbf{r}}}{i\mathbf{k}}}
        {\omega - (\omega_{j\mathbf{k}} - \omega_{i\mathbf{k}}) + i \mathrm{sgn}(\omega_{j\mathbf{k}} - \omega_{i\mathbf{k}}) \eta} \\ 
        \epsilon_{\mathbf{G}0}(\omega, \mathbf{q} \rightarrow 0) &=    
        \frac{- 4\pi \alpha \: i}{V_\mathrm{cell} m_e |\mathbf{G}|} 
        \sum_{ij\mathbf{k}} \frac{f_{i\mathbf{k}} - f_{j\mathbf{k}}}{\omega_{j\mathbf{k}} - \omega_{i\mathbf{k}}}
        \lim_{\eta \rightarrow 0} 
        \frac{\mel{i\mathbf{k}}{e^{-i \mathbf{G}\cdot\mathbf{r}}}{j\mathbf{k}} \: \Hat{q} \cdot \mel{j\mathbf{k}}{\nabla}{i\mathbf{k}}}
        {\omega - (\omega_{j\mathbf{k}} - \omega_{i\mathbf{k}}) + i \mathrm{sgn}(\omega_{j\mathbf{k}} - \omega_{i\mathbf{k}}) \eta}
    \end{align}
    \end{widetext}

\section{Comparing to DarkELF} \label{sec:DarkELF_comparison}

    \begin{figure*}[ht]
        \includegraphics[width=1\linewidth]{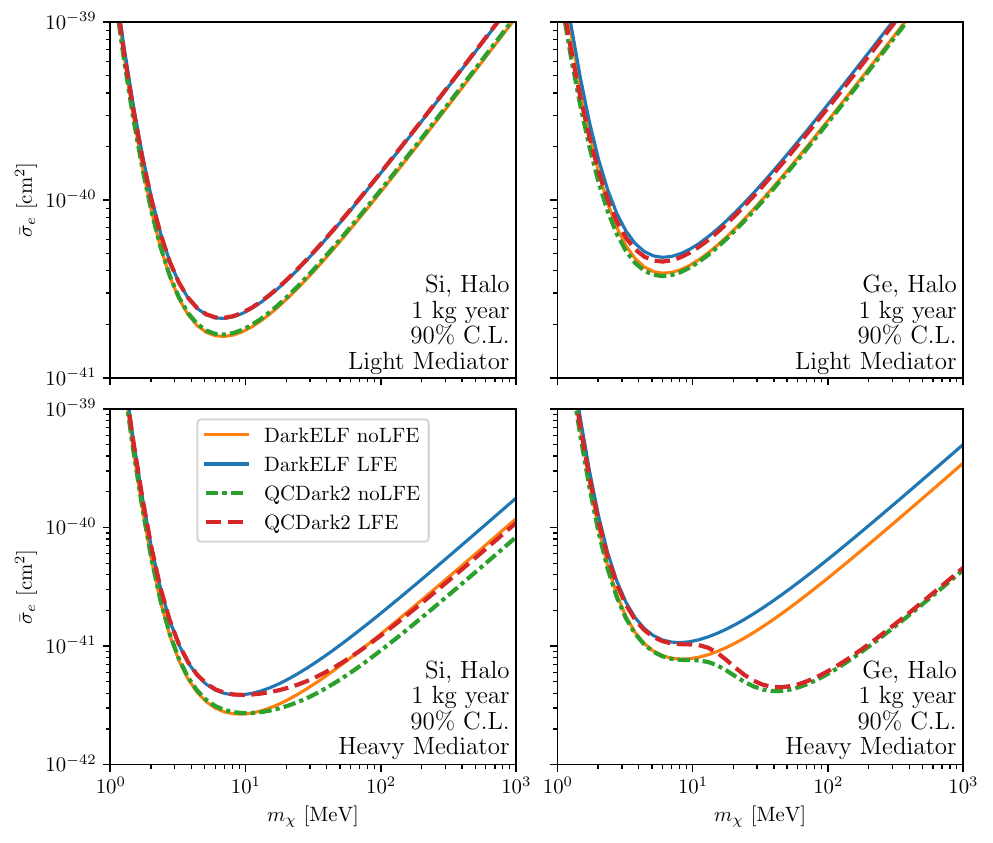}
        \caption{Projections for the 90\% C.L.~sensitivity in Si and Ge, assuming no backgrounds and a one-electron threshold. We compare the RPA dielectric function calculated with \textsc{GPAW} from~\cite{DarkELF} with and without LFEs to the RPA dielectric function calculated with \textsc{QCDark2}.}
        \label{fig:projection_DarkELF}
    \end{figure*}

    The projected sensitivities calculated from the \textsc{GPAW} dielectric functions tabulated as part of the \textsc{DarkELF} code are compared to those calculated with \textsc{QCDark2} in Fig.~\ref{fig:projection_DarkELF}. Both codes provide RPA dielectric functions with and without LFEs, however \textsc{DarkELF} only includes momentum transfers up to 6 $\alpha m_e$ compared to 25 $\alpha m_e$ for Si or 20 $\alpha m_e$ for Ge in \textsc{QCDark2}. In the light mediator limit, the DM form factor contributes an extra factor of $q^{-4}$, which suppresses high-momentum contributions. Therefore the \textsc{DarkELF} and \textsc{QCDark2} projections with and without LFEs agree very well for a light mediator. 

    In the heavy mediator limit, the high-momentum contributions that become kinematically relevant for masses larger than 10~MeV are not suppressed by the form factor, so the projected sensitivities calculated with \textsc{QCDark2} and \textsc{DarkELF} start to diverge at this mass due to the absence of data above 6 $\alpha m_e$ in the \textsc{DarkELF} dielectric function (see also the electron recoil spectrum at $m_{\chi} = 1$ GeV in Fig.~\ref{fig:recoil_spectrum_codes}). The high-momentum contributions are especially important for Ge due to transitions from the $3d$ orbitals, which \textsc{DarkELF} freezes in the pseudopotential core as discussed in Section~\ref{sec:Results:Reach}.

\bibliography{bib.bib}

\end{document}